\documentclass[superscriptaddress,groupedaddress,nofootnoteinbib,11pt]{article}
\pdfoutput=1
\usepackage{jheppub} % for details on the use of the package, please see the JINST-author-manual

\usepackage[T1]{fontenc}
\usepackage{amsmath,amsfonts,amssymb,amsthm,mathrsfs,bm}
\usepackage{xcolor}
\usepackage{graphicx} \graphicspath{{figures_paper}}
\usepackage{caption,subcaption}
\usepackage{hyperref} \hypersetup{}
\usepackage{orcidlink}

\usepackage{ximpleinput}
\newcommand{\ind}[1]{^{#1}{}}
\newcommand{\mbar}{\bar{m}_P}
\newcommand{\ren}{\text{ren}}
\allowdisplaybreaks[3]

\makeatletter
\def\@fpheader{~}
\makeatother

\title{Quantum fields in boson star spacetime}

\author[a]{Paul M. Saffin \orcidlink{0000-0002-4290-3377}}
\author[a]{and Qi-Xin Xie \orcidlink{0009-0000-2391-863X}}
\affiliation[a]{School of Physics and Astronomy, University Park, University of Nottingham,\\ Nottingham NG7 2RD, United Kingdom}

\emailAdd{paul.saffin@nottingham.ac.uk}
\emailAdd{qixin.xie@nottingham.ac.uk}
\abstract{Boson stars have been extensively studied in classical gravity, but their quantum properties remain comparatively unexplored. In this paper, we compute the quantum scalar fields and stress tensor in boson star spacetimes within the framework of semiclassical gravity. Divergences are regularized using Pauli-Villars fields, and accurate numerical results are obtained through spectral methods. Employing coherent states enables a direct comparison between the classical part of the stress tensor and the quantum fluctuation. Our results indicate that strong spacetime curvature is the primary source of large quantum effects. The renormalized quantum energy density is mostly positive but the radial pressure is negative, suggesting that classical boson star solutions require modification once quantum effects are included. Moreover, in regimes of large curvature, the quantum fluctuations can constitute a significant fraction of the total stress tensor. The methods developed here can be generalized to other compact objects and used to study their response to quantum corrections.}

\begin{document}
\maketitle
\flushbottom

\section{Introduction}
\label{sec:introduction}

General relativity provides an accurate description of the large-scale structure of spacetime, while quantum field theory describes matter with unprecedented precision. It is appealing to combine these two theories and investigate how spacetime geometry and quantum effects affect one another. Semiclassical gravity is such an attempt. It treats the metric as a classical field while matter is treated quantum mechanically; the metric affects matter through its covariant equation of motion, and matter backreacts to the metric through the expectation value of its quantum stress tensor in Einstein's equation. Although not a quantum gravity theory, semiclassical gravity is well suited for studying quantum effects in strong gravitational environments away from gravitational singularities, where fully quantized gravity is not yet required \cite{Birrell:1982ix,Fulling:1989nb,Wald:1995yp,Mukhanov:2007zz,Parker:2009uva,Hu:2020luk}. 

Studying quantum effects in curved spacetime has produced fruitful achievements. One example is cosmological particle creation \cite{Parker:1968mv}, where the expansion of the universe generates particles from the vacuum. The resulting vacuum fluctuations \cite{Bunch:1978yq} serve as the seeds for the structure of the present universe. Another example is Hawking radiation \cite{Hawking:1975vcx}, which predicts that black holes can evaporate by emitting particles. This gives rise to the information loss paradox \cite{Hawking:1976ra}, raising the question of whether quantum information falling into a black hole is destroyed or preserved. 

Beyond these prestigious topics, semiclassical gravity has been applied to study fluid stars \cite{Carballo-Rubio:2017tlh}. These fluid stars are found to be quite different from their counterparts in fully classical theory \cite{Arrechea:2021pvg,Arrechea:2021xkp,Reyes:2023fde,Arrechea:2023oax,Melella:2025tkh}, which exhibit negative energy densities within their surfaces and compactness exceeding the Buchdahl limit \cite{Buchdahl:1959zz}. This research shows that semiclassical effects can significantly change stellar solutions and reveal more exotic properties. Semiclassical gravity has also been used to study gravitational collapse \cite{Brady:1998fh,Vachaspati:2007hr,Greenwood:2008ht,Vachaspati:2018pps,Benitez:2020szx,Berczi:2020nqy,Guenther:2020kro,Berczi:2021hdh,Berczi:2024yhb,Tomasevic:2025clf}, quantum superradiance \cite{Balakumar:2022yvx,Balakumar:2020gli,Breen:2024ggu,Montagnon:2025vtk,Alvarez-Dominguez:2024ahv,Fu:2025ztk}, oscillating stars \cite{delRio:2026exc} and many other topics.

Similarly to fluid stars, boson stars are another class of spherically symmetric, gravitationally bound objects, which are made entirely of bosonic fields \cite{Kaup:1968zz}. Depending on the field mass and its couplings, boson stars can reach macroscopic mass and size, and exhibit large compactness. They have broad astrophysical relevance, including as dark matter candidates \cite{Chavanis:2011zi,Rindler-Daller:2011afd,Eby:2015hsq,Braaten:2019knj,Chen:2020cef} and as compact mimickers to black holes \cite{Macedo:2013jja,Vincent:2015xta,Sennett:2017etc,Mark:2017dnq,Cardoso:2019rvt}. The gravitational waves they radiate can be used to constrain the mass of new particles \cite{CalderonBustillo:2020fyi}. Recent studies show that they can form composite structures \cite{Copeland:2014qra,Xie:2021glp,Jaramillo:2024smx,Jaramillo:2024cus} and superradiate scalar waves \cite{Saffin:2022tub,Gao:2023gof,Chang:2024xjp}.

Although the majority of research studies boson stars in a classical context, their quantum properties have been continuously investigated since their discovery \cite{Ruffini:1969qy}. They were first studied in the ground state \cite{Breit:1983nr,Ho:2002vz,Barranco:2010ib}, and then in the combination of the ground state and the first few radially excited states \cite{Matos:2007zza,Bernal:2009zy,Urena-Lopez:2010zva}, which are shown to be stable and better reproduce galactic rotation curves. \cite{Alcubierre:2022rgp} generalized these works and found several examples of the multi-$\ell$ multi-state boson stars in semiclassical gravity. However, in addition to that these works mainly focus on theory with a single real scalar field and states with definite particle numbers, which do not resemble classical dynamics very well, the regularization method they use to compute stress tensor is normal ordering, which breaks diffeomorphism invariance in curved spacetime\footnotemark{}. Consequently, the quantum fluctuation energy is dropped, which has an obvious influence in curved spacetime, as we will demonstrate later. 
\footnotetext{Recently, \cite{Javed:2025ogr} studied semiclassical boson stars using a Hamiltonian version of the semiclassical Einstein equation. Rather than specifying a quantum state, the expectation values of the scalar field are specified to denote a bounded configuration. }

In this paper, we study the quantum effects in boson stars using improved methods. Computing the renormalized stress tensor in curved spacetime is generally challenging, because infinitely many quantum mode functions are involved, which typically lack analytic expressions. We use the Pauli-Villars regularization method \cite{Pauli:1949zm}. By introducing additional ghost fields with specific masses and opposite energy contribution into the Lagrangian, the divergences in large momentum are regularized. The resultant quartic and quadratic divergences from the large masses of the ghost fields are then absorbed into the cosmological constant and Newton's gravitational constant, leaving behind only logarithmic divergences. 

The diffeomorphism invariance is manifestly preserved as a result of the Lagrangian being a Lorentz scalar. Moreover, this method is suitable for numerical calculation, because the regularization is implemented at the level of summing each set of modes, before the integration of momentum is performed. It can be readily generalized to other curved spacetimes, including dynamical ones. The Pauli-Villars method has been used in computing stress tensor \cite{Bernard:1977pq,Mera:2014bza}, black hole entropy \cite{Demers:1995dq}, cosmological correlations \cite{Weinberg:2010wq}, and so on in curved spacetimes. In particular, it is used in simulating gravitational collapse \cite{Berczi:2020nqy,Berczi:2021hdh,Berczi:2024yhb}, where the Choptuik scaling law \cite{Choptuik:1992jv} of critical collapse is compared in the classical and quantum cases. 

Another improvement in our paper is the use of coherent states, which are eigenstates of the annihilation operator and minimize the uncertainty relation. In such states, the stress tensor can be naturally split into a classical part and a quantum fluctuation. The classical part follows the form of the stress tensor of a classical boson star, while the quantum fluctuation provides corrections to it; this decomposition provides a good basis for studying the differences between classical and quantum solutions. For a comprehensive review of the properties of coherent states, see \cite{Zhang:1990fy}.

The stress tensor is obtained numerically. We divide this problem into two stages: the metric is computed by solving the classical Einstein equation using the Newton-Raphson method \cite{press2007numerical}, while the quantum fields are solved as a linear eigenproblem, both are solved on lattices. We adopt spectral methods to approximate the derivatives \cite{trefethen2000spectral,boyd2001chebyshev}, whose numerical error has an exponential decrease as the number of grid points increases. Spectral methods have seen rapid adoption in numerical relativity during recent years; see \cite{Grandclement:2007sb} for a review.

The paper is organized as follows. In \sect{sec:model}, we first introduce how to quantize the field in curved static spacetime. Next we introduce the boson star solution of the classical Einstein equation. We then apply our model to coherent states, and focus on quantum fields in a spherically symmetric spacetime. Finally, we describe the regularization and renormalization scheme we use, where the regularization conditions are derived in Appendix~\ref{sec:PVinFRW}. In \sect{sec:results}, we first introduce the numerical setup, which extensively uses the spectral method described in Appendix~\ref{sec:spectral}. We then solve the boson star metrics, in which we compute quantum mode functions. Afterwards, we evaluate the stress tensor by computing the contribution of these mode functions. The relationship between the stress tensor and curvature, and the proportion of the quantum fluctuation within the stress tensor are studied. In \sect{sec:summary}, we summarize our results. Finally, we discuss the backreaction to spacetime and extensions of our method to other compact objects. We use natural units $\hbar=c=1$ throughout this paper. When denoting spacetime indices, Greek letters run from $0$ to $3$ and Latin letters run from $1$ to $3$. 

\section{Model}
\label{sec:model}

In the following, we review field quantization in static curved spacetimes (see e.g. \cite{Birrell:1982ix,Fulling:1989nb,Wald:1995yp,Mukhanov:2007zz,Parker:2009uva,Hu:2020luk}). Next, we introduce boson stars as classical solutions. Then, we construct coherent states adapted to the symmetry of spherically symmetric spacetimes, and discuss quantum fields in these backgrounds. Finally, we specify the regularization and renormalization scheme we use. 

\subsection{Quantization in curved spacetime}
\label{sec:quantization}

Canonical quantization proceeds analogously to the Minkowski case, except that the metric is now generic. The gravitational field is treated as a classical field. For a static spacetime, we adopt the following form of the line element 
\begin{equation}
    ds^2=-\ai^2(\bm{x})\d t^2+\gi_{ij}(\bm{x})\d x^i\d x^j,
    \label{eq:staticg}
\end{equation}
where spherical coordinates $\{t,\bm{x}\}=\{t,r,\qi,\cphi\}$ are used for later convenience. This spacetime can be foliated by a set of constant-$t$ spacelike hypersurfaces $\Si$, whose future-directed unit normal vector is $n^{\mu}=(\ai^{-1},\bm{0})$. The lapse function $\ai$ and the induced metric $\gi_{\mn}=g_{\mn}+n_{\mu}n_{\nu}$ are functions of the spatial coordinates $\bm{x}$ only, and the metric determinants are related by $\sqrt{-g}=\ai\sqrt{\gi}$. 

The matter fields, by contrast, are quantum-mechanical and represented by operators. We consider a set of real scalar fields $\phi=\{\phi\ind{n}\},n=1,2,\dots,2N$, with the same mass $\mu$, whose Lagrangian density is 
\begin{align}
    \mcl{L}_M=\sqrt{-g}\lt(-\f12g^{\mn}\n_{\mu}\phi\cdot\n_{\nu}\phi-\f12\mu^2\phi\cdot\phi\rt),
    \label{eq:Lagrangian}
\end{align}
where $\n_{\mu}$ is covariant derivative and $\phi\cdot\phi=\phi\ind{m}\phi\ind{n}\di_{mn}$. 
The conjugate momentum is defined as 
\begin{equation}
    \pi\ind{n}=\sqrt{\gi}n^{\mu}\n_{\mu}\phi\ind{n}.
\end{equation}
Field quantization is performed in each hypersurface $\Si$ and the Hamiltonian takes the form
\begin{align}
    H &= \int_{\Si} d\Si\sqrt{\gi}\ai\lt(\f12\f{\pi\cdot\pi}{\gi}+\f12\gi^{ij}\n_i\phi\cdot\n_j\phi+\f12\mu^2\phi\cdot\phi\rt).
\end{align}
By imposing the canonical commutation relations in each hypersurface 
\begin{equation}
    [\phi\ind{m}(t,\bm{x}),\pi\ind{n}(t,\bm{y})]=i\di^3(\bm{x}-\bm{y})\di^{mn},~~
    [\phi\ind{m}(t,\bm{x}),\phi\ind{n}(t,\bm{y})]=[\pi\ind{m}(t,\bm{x}),\pi\ind{n}(t,\bm{y})]=0
    \label{eq:commutationPhiPi},
\end{equation}
we obtain the equation of motion for the scalar fields
\begin{equation}
    -\Box\phi\ind{n}+\mu^2\phi\ind{n}=0.
    \label{eq:KGeq}
\end{equation}

For solutions of the Klein-Gordon equation \eref{eq:KGeq}, we define the scalar product 
\begin{equation}
    (f_I,f_J)=-i\int_{\Si}d\Si\sqrt{\gi}n^{\mu}\lt(f_I\n_{\mu}f_J^*-f_J^*\n_{\mu}f_I\rt),
    \label{eq:scalarProduct}
\end{equation}
where the subscripts $I,J$ label different solutions. Although the spacetime is curved, this equation still admits a set of solutions associated to each momentum mode, which can form an orthogonal and complete basis\footnote{Explicit examples in FRW spacetime can be found in Appendix~\ref{sec:PVinFRW}}, analogous to plane waves in Minkowski spacetime. We therefore use the subscripts $I,J$ to denote momentum modes, which become a set of quantum numbers like those of the hydrogen atom, and define positive-frequency modes 
\begin{equation}
    f_I\ind{n}=e^{-i\oi_I\ind{n}t}f_I\ind{n}(\bm{x}),~\oi_I\ind{n}>0; 
\end{equation}
thus $f_I\ind{n}^*$ with a positive sign in the exponent are negative-frequency modes. The orthogonality relations among these modes are 
\begin{equation}
    (f_I\ind{n},f_J\ind{n})=\di_{IJ},~~(f_I\ind{n}^*,f_J\ind{n}^*)=-\di_{IJ},~~(f_I\ind{n},f_J\ind{n}^*)=0.
    \label{eq:orthonormality}
\end{equation}
For non-zero mass $\mu\ne0$, the mode functions can be either bound states with $\oi_I\ind{n}<\mu$ or scattering states with $\oi_I\ind{n}>\mu$. 

Using the set of complete mode functions, we can expand the scalar fields 
\begin{equation}
    \phi\ind{n} = \sum_I A_I\ind{n}f_I\ind{n}+A_I\ind{n}^{\dag}f_I\ind{n}^*,
    \label{eq:modeExpansion}
\end{equation}
where the coefficients are the annihilation operator $A_I\ind{n}$ and the creation operator $A_I\ind{n}^\dag$. These two operators are time-independent. They satisfy the following commutation relations
\begin{equation}
    [A_I\ind{m},A_J\ind{n}^{\dag}]=\di_{IJ}\di^{mn},~~[A_I\ind{m},A_I\ind{n}]=[A_J\ind{m}^{\dag},A_J\ind{n}^{\dag}]=0,
    \label{eq:commutationaadagger}
\end{equation}
which can be shown equivalent to the commutation relations in \eref{eq:commutationPhiPi}. The vacuum $|0\rangle$ is defined by 
\begin{equation}
    A_I\ind{n}|0\rangle=0,
\end{equation}
from which all states in the Hilbert space can be constructed. Although we use the Kronecker delta here, it is easy to consider continuum labels, say $\bm{k}$, by changing the Kronecker delta $\di_{IJ}$ to the Dirac delta $\di^3(\bm{k}_I-\bm{k}_J)$ and the summation $\sum_I$ to an integration $\int d\bm{k}$ in these and subsequent equations. 

Semiclassical gravity combines the classical geometric part and the quantum matter part to obtain the action 
\begin{equation}
    S=\int d^4 x\sqrt{-g}\lt(\mbar^2\f{R}{2}-\Li+\mcl{O}(R^2,R^{\mn}R_{\mn},R^{\mn\rs}R_{\mn\rs})\rt)+\mcl{L}_M,
    \label{eq:action}
\end{equation}
where $R$ is the Ricci scalar, $\mbar=m_P/\sqrt{8\pi}$ is the reduced Planck mass, $m_P=1/\sqrt{G}$ is the Planck mass, and $G$ is Newton's constant. In the above equation, we include the cosmological constant $\Li$ and higher-order curvature terms. They are necessary to absorb divergences in quantum expectation values, and their renormalized values match observational values, as we will show later in \sect{sec:PVregularization}. In this paper, we ignore the higher-order curvature terms by choosing the renormalization scale where their contributions are vanishing. Their dependence on the running of the renormalization scale is only logarithmic, which is subleading compared to those in $\Li$ and $\mbar^2$, and thus is also ignored. 

Varying the action \eref{eq:action}, we obtain the semiclassical Einstein equation
\begin{equation}
    \mbar^2G_{\mn}+\Li g_{\mn}=\langle T_{\mn} \rangle ,
    \label{eq:semiEinstein}
\end{equation}
where the quantum stress tensor is 
\begin{equation}
    T_{\mn}=\n_{\mu}\phi\cdot\n_{\nu}\phi-\f12g_{\mn}\lt(\n^{\li}\phi\cdot\n_{\li}\phi+\mu^2\phi\cdot\phi\rt), 
    \label{eq:stressTensor}
\end{equation}
and $\langle T_{\mn} \rangle$ denotes its expectation value in a quantum state. Formally \eref{eq:semiEinstein} has the same structure as the classical Einstein equation, but with the classical stress tensor replaced by the expectation value of the quantum stress tensor. The dependence on quantum states of the geometric part on the left-hand side arises from the matter part on the right-hand side, which can vary substantially in different states. Quantum corrections to the metric are expected to be large in strong gravitational environments, such as near black holes and close to the big bang singularity.  

\subsection{Classical boson star}
\label{sec:classicalBS}

Before continuing with the discussion of quantum fields, let us first consider the classical boson star. This is a self-gravitating localized object composed of bosons, which is regular everywhere, and the spacetime is static and asymptotically flat. Such solutions typically possess spherical symmetry for stability, so the line element can be in general written as 
\begin{equation}
    ds^2 = -\f{1}{b(r)^2s(r)^2}\d t^2+b(r)^2\d r^2+r^2d\Oi^2,
    \label{eq:sphericalg}
\end{equation}
where the metric functions $b,s$ depend on the radial coordinate $r$ only, and \mbox{$d\Oi^2=\d\qi^2+\sin^2\qi\,\d\cphi^2$}. $b$ and $s$ are regular functions and never vanish; asymptotic flatness requires them to approach unity for large distances, while the absence of a horizon implies that neither diverges anywhere. The stress tensor satisfying the Einstein equation is therefore diagonal but can be anisotropic, and the boson star is an explicit example as we will see.

For the simplest boson star composed of a complex scalar field \cite{Kaup:1968zz}, which is equivalent to the Lagrangian \eref{eq:Lagrangian} with two real scalars, no self-interaction terms are present; the boson star is condensed by gravitational attraction. The scalar fields satisfy the same Klein-Gordon equation as in \eref{eq:KGeq}, and the stress tensor in the Einstein equation \eref{eq:semiEinstein} consists solely of classical fields. The two fields evolve periodically
\begin{equation}
    \phi\ind{0r}=\si(r)\cos(\oi t),~~\phi\ind{0i}=-\si(r)\sin(\oi t),
    \label{eq:classicalEvolution}
\end{equation}
where $\oi$ is the frequency and $\si$ is a real radial function, and we rename the field indices for later convenience. 

The classical stress tensor $T_{\mn}^{(\text{c})}$ has the same form as in \eref{eq:stressTensor}, but evaluated for classical fields. Using the separation of variables in \eref{eq:classicalEvolution}, the energy density $\ri^{(\text{c})}$, radial pressure $p_r^{(\text{c})}$ and tangent pressure $p_t^{(\text{c})}$ can be written as 
\begin{equation}
\begin{aligned}
    \ri^{(\text{c})}&=-T_0{}^0{}^{(\text{c})}=\f{b^2s^2}{2}\oi^2\si^2+\f{1}{2b^2}(\si')^2+\f12\mu^2\si^2, \\
    p_r^{(\text{c})}&=T_1{}^1{}^{(\text{c})}=\f{b^2s^2}{2}\oi^2\si^2+\f{1}{2b^2}(\si')^2-\f12\mu^2\si^2, \\
    p_t^{(\text{c})}&=T_2{}^2{}^{(\text{c})}=T_3{}^3{}^{(\text{c})}=\f{b^2s^2}{2}\oi^2\si^2-\f{1}{2b^2}(\si')^2-\f12\mu^2\si^2,
\end{aligned}
\end{equation}
where $'$ denotes differentiation with respect to $r$. All off-diagonal components vanish. Thus $T_{\mn}^{(\text{c})}$ is manifestly static and spherically symmetric, consistent with the symmetry of the spacetime. Additionally, the pressures differ in different directions and vary with position, implying that the spacetime is neither homogeneous nor isotropic, similar to the Schwarzschild black hole. 

The equations of motion become 
\begin{equation}
\begin{aligned}
    b'&=\f{b-b^3}{2r}+\f{b^3r}{2\mbar^2}\lt(-T_0{}^0{}^{(\text{c})}+\Li\rt),\\
    s'&=-\f{sb^2r}{2\mbar^2}\lt(-T_0{}^0{}^{(\text{c})}+T_1{}^1{}^{(\text{c})}\rt),\\
    \si''&+\lt(\f2r-\f{2b'}{b}-\f{s'}{s}\rt)\si'+b^2\lt(b^2s^2\oi^2-\mu^2\rt)\si=0,
    \label{eq:cBSeq}
\end{aligned}
\end{equation}
where the cosmological constant is set to $\Li=0$. Assuming a series solution near the origin, regularity requires $b\approx 1+b_2r^2+\dots$, $s\approx s_0+s_2r^2+\dots$, and $\si\approx \si_0+\si_2r^2+\dots$, where the coefficients in the expansions are constants. Asymptotic flatness demands $b(r\to\infty)=s(r\to\infty)=1$, so $\si$ vanishes at infinity. In short, the four boundary conditions for \eref{eq:cBSeq} can be chosen as 
\begin{equation}
    b(r=0)=1,~~\si(r\to\infty)=0,~~\si'(r=0)=0,
    \label{eq:classicalBC}
\end{equation}
with either $s(r=0)$ or $\si(r=0)$ left as a free parameter to determine the final boundary condition. 

The symmetries in \eref{eq:cBSeq} allow the mass $\mu$ to be absorbed into dimensional quantities
\begin{equation}
    \t{r}=r\mu,~~\t{\oi}=\t{\oi}/\mu,~~\t{\si}=\si/\mu,~~\t{m}_P=\mbar/\mu,
    \label{eq:dimensionless}
\end{equation}
where $\sim$ denotes dimensionless quantities. Time-translation invariance similarly allows the frequency $\oi$ to be absorbed 
\begin{equation}
    \h{s}=\t{s}\t{\oi}.
    \label{eq:hats}
\end{equation}
Equivalently, we can set $\mu=\oi=1$ in \eref{eq:cBSeq}. 
Lastly, the Planck mass can be absorbed into the field
\begin{equation}
    \h{\si}=\t{\si}/\t{m}_P,
    \label{eq:hatsigma}
\end{equation}
which implies that the metric functions $b,s$ are invariant under changes in $\mbar$. These variables will be useful when we compute numerical solutions later in \sect{sec:results}. 

For boson stars, key physical quantities include the mass, radius, compactness, and curvature. To obtain the mass, we can write the metric function the following way to define the Misner-Sharp mass function $m(r)$ \cite{Misner:1964je}
\begin{equation}
    b(r)^{-2} = 1-\f{2Gm(r)}{r},\footnotemark
\end{equation}
\footnotetext{Some static and spherically symmetric metrics takes a similar form to this, e.g., the Bardeen black hole \cite{bardeen1968non}. However, for the boson star $m(r)$ is chosen such that no horizon exists, and the metric components satisfy $g_{tt}\ne g_{rr}^{-1}$. }
and the total mass is then 
\begin{equation}
    M = \lim_{r\to\infty}m(r) = \int_0^{\infty}4\pi r^2\ri^{(\text{c})}dr.
    \label{eq:BSmass}
\end{equation}
To obtain the second equality, the Einstein equation is used. One may also define the Komar mass using the timelike Killing vector $K^{\mu}=(1,\bm0)$; for static spacetimes it equals to the ADM mass, and both coincide with \eref{eq:BSmass}. 

Boson stars do not possess well located surfaces; typically the radius is defined to be where most of the mass is included. For example,
\begin{equation}
    0.99M = \int_0^{R_{99}}4\pi r^2\ri^{(\text{c})}dr
    \label{eq:BSradius}
\end{equation}
defines the radius $R_{99}$ where $99\%$ of the total mass is included. The compactness is the ratio of the mass to the radius 
\begin{equation}
    C = \f{GM}{R_{99}}.
    \label{eq:BScompactness}
\end{equation}
Buchdahl's theorem \cite{Buchdahl:1959zz} states that the compactness of a static, spherically symmetric matter configuration has an upper limit $GM/R_{99}<4/9$. The curvature of a classical boson star can be represented by the trace of the stress tensor
\begin{equation}
    R = \f{-T}{\mbar^2} = \f{1}{\mbar^2}\lt(-b^2s^2\oi^2\si^2+\f{(\si')^2}{b^2}+2\mu^2\si^2\rt),
    \label{eq:BSricci}
\end{equation}
where the first equality follows from the Einstein equation. 

\subsection{Coherent state}
\label{sec:coherentState}

In this section, we construct coherent states whose expectation values of the stress tensor are static and spherically symmetric. Therefore, the mode index can be represented by three numbers $I=(k,l,m)$. For brevity, we temporarily omit the field indices. 

In a free field theory, each momentum mode behaves as a harmonic oscillator. For a given mode $I$, a coherent state is obtained by displacing the vacuum  
\begin{equation}
    |\chi_I\rangle=D_I(\chi_I)|0\rangle,
\end{equation}
where 
\begin{equation}
    D_I(\chi_I)=\exp(\chi_IA_I^{\dag}-\chi_I^*A_I)
\end{equation}
is the displacement operator, and $\chi_I$ is a complex parameter designating the state. 

It is easily seen that $D_I$ is a unitary operator 
\begin{equation}
    D_I^\dag(\chi_I)=D_I^{-1}(\chi_I),
\end{equation}
and can be shown to satisfy the following relations \cite{Caves:1981hw}
\begin{equation}
\begin{aligned}
    D_I^\dag(\chi_I)A_I D_I(\chi_I) &= A_I+\chi_I, \\
    D_I^\dag(\chi_I)A_I^\dag D_I(\chi_I) &= A_I^\dag+\chi_I^*.
    \label{eq:displacement}
\end{aligned}
\end{equation}
Using these relations, it can be shown that the coherent state is an eigenstate of the annihilation operator with eigenvalue $\chi_I$
\begin{equation}
    A_I|\chi_I\rangle=\chi_I|\chi_I\rangle.
\end{equation}
In particular, the vacuum state $|0\rangle$ is a coherent state with eigenvalue $0$. The coherent state also saturates the uncertainty relation between the field and its conjugate momentum so that the quantum corrections are minimized.  

When all modes are included, the full displacement operator is 
\begin{equation}
    D(\chi) = D_{I_1}(\chi_{I_1})D_{I_2}(\chi_{I_2})\dots.
\end{equation}
Because operators associated with different modes commute, the ordering on the right-hand side of the equation can be arbitrary. The corresponding coherent state is therefore 
\begin{equation}
    |\chi\rangle = |\chi_{I_1}\chi_{I_2}\dots\rangle = D(\chi)|0\rangle.
\end{equation}

The expectation value of the stress tensor \eref{eq:stressTensor} in this coherent state is 
\begin{align}
    \langle\chi|T_{\mn}|\chi\rangle &= \sum_{I,J} \mcl{T}_{\mn}(f_I,f_J)\langle\chi|A_IA_J|\chi\rangle + \mcl{T}_{\mn}(f_I,f_J^*)\langle\chi|A_IA_J^\dag|\chi\rangle \notag \\ & ~~~~ + \mcl{T}_{\mn}(f_I^*,f_J)\langle\chi|A_I^\dag A_J|\chi\rangle + \mcl{T}_{\mn}(f_I^*,f_J^*)\langle\chi|A_I^\dag A_J^\dag|\chi\rangle, \\ 
    &= \n_{\mu}\cphi\cdot\n_{\nu}\cphi-\f12g_{\mn}\lt(\n^{\li}\cphi\cdot\n_{\li}\cphi+\mu^2\cphi\cdot\cphi\rt) + \sum_I\mcl{T}_{\mn}(f_I,f_I^*),
    \label{eq:coherentStressTensor}
\end{align}
where 
\begin{equation}
    \mcl{T}_{\mn}(f_I,f_J)=\n_{\mu}f_I\cdot\n_{\nu}f_J-\f12g_{\mn}\lt(\n^{\li}f_I\cdot\n_{\li}f_J+\mu^2f_I\cdot f_J\rt),
\end{equation}
and 
\begin{equation}
    \cphi = \sum_I \chi_If_I+\chi_I^*f_I^*
\end{equation}
is a classical field. The last term in \eref{eq:coherentStressTensor} is the quantum fluctuation which does not depend on the quantum states; the remaining terms of the equation take precisely the form of the classical stress tensor. As the expectation value of the field is 
\begin{equation}
    \langle\chi|\phi|\chi\rangle = \sum_I \chi_If_I+\chi_I^*f_I^*,
\end{equation}
it follows that $\cphi$ is the mean field and the expectation value of the stress tensor splits into the classical stress tensor from the classical field $\cphi$ plus the quantum fluctuation. %This gives us the result in \eref{eq:stressCQ}. 

To ensure that the stress tensor expectation value is static and spherically symmetric, the coherent state must satisfy additional conditions, which fix the form of $\cphi$. A simple choice is to choose $\cphi\ind{1}=\cphi(r)\cos(\oi t),\cphi\ind{2}=-\cphi(r)\sin(\oi t)$, consistent with a boson star solution. $\chi_I$ are then obtained using the orthonormality relation in \eref{eq:orthonormality}
\begin{equation}
    \chi_I=(\cphi,f_I),~~\chi_I^*=-(\cphi,f_I^*).
\end{equation}
For this choice, it is directly seen that if $m\ne0$ or $l\ne0$, $\chi_{klm}\propto\int Y^{lm*}(\qi,\cphi)\sin(\qi)d\qi d\phi=0$, so only the radial modes are excited. The staticity and spherical symmetry of the quantum fluctuation will be proved in \sect{sec:sphere}. 

\subsection{Spherically symmetric spacetime}
\label{sec:sphere}

We now return to the discussion of quantum fields. From this point onward, we focus on spherically symmetric and static spacetimes whose metric is given by \eref{eq:sphericalg}. Owing to the symmetries of this spacetime, the mode functions can be separated as 
\begin{equation}
    f_I\ind{n} = \f{1}{\sqrt{2\oi_{kl}\ind{n}}}e^{-i\oi_{kl}\ind{n}t}
    \f{v_{kl}\ind{n}(r)}{r}Y^{lm}(\qi,\cphi),
\end{equation}
where the quantum numbers are $I=(k,l,m)$ and $Y^{lm}$ are spherical harmonics. The orthonormality relations in \eref{eq:orthonormality} then give 
\begin{equation}
    \int dr\, b^2s v_{k_1l}\ind{n}(v_{k_2l}\ind{n})^*=\di_{k_1k_2},
    \label{eq:normalization}
\end{equation}
where we have used 
\begin{equation}
    \int d\cphi\int d\qi\sin\qi ~Y^{l_1m_1*}(\qi,\cphi)Y^{l_2m_2}(\qi,\cphi)=\di_{l_1l_2}\di_{m_1m_2}.
    \label{eq:orthonormalityYlm}
\end{equation}
From the equation of motion \eref{eq:scalarEq} we derive later, $v_{kl}\ind{n}$ can be chosen to be real. 

To compare the quantum case with the classical solutions, we consider two fields $\{\phi^{0r},\phi^{0i}\}$. The coherent state, which minimizes the quantum uncertainty relation and closely reproduces classical evolution, provides a natural framework for this comparison. We have constructed the coherent state $|\chi\rangle$ adapted to the symmetries of the spacetime in \sect{sec:coherentState}. By design, its field expectation values match the classical behaviour in \eref{eq:classicalEvolution}
\begin{equation}
    \langle\chi|\phi\ind{0r}|\chi\rangle=\si(r)\cos(\oi t),~~\langle\chi|\phi\ind{0i}|\chi\rangle=-\si(r)\sin(\oi t).
    \label{eq:classicalMean}
\end{equation}

When expressed in terms of the mode functions, the stress tensor naturally splits into a classical part and a quantum fluctuation
\begin{equation}
    \langle\chi|T_{\mn}|\chi\rangle = T_{\mn}^{(\text{c})}+T_{\mn}^{(\text{q})},
    \label{eq:stressCQ}
\end{equation}
where $T_{\mn}^{(\text{c})}$ has exactly the same form as in the classical solution, so it is static and spherically symmetric, and 
\begin{equation}
    T_{\mn}^{(\text{q})} = \langle0|T_{\mn}|0\rangle = \sum_I \n_{\mu}f_I\cdot\n_{\nu}f_I^*-\f12g_{\mn}\lt(\n^{\li}f_I\cdot\n_{\li}f_I^*+\mu^2f_I\cdot f_I^*\rt)
\end{equation}
is the quantum fluctuation. We define the quantum energy density $\ri^{\text{(q)}}$ and radial pressure $p_r^{\text{(q)}}$ accordingly
\begin{equation}
    \ri^{(\text{q})}=-T_0{}^0{}^{(\text{q})},~~
    p_r^{(\text{q})}=T_1{}^1{}^{(\text{q})}.
    \label{eq:quantumDensity}
\end{equation}
A direct calculation shows that the quantum contribution is also static and spherically symmetric. For instance, the energy density is 
\begin{align}
    \ri^{(\text{q})}&=\sum_n\f{1}{4\pi}\sum_{kl}\f{2l+1}{2\oi_{kl}\ind{n}}\Bigg(\f12b^2s^2(\oi_{kl}\ind{n})^2\f{|v_{kl}\ind{n}|^2}{r^2}+\f{1}{2}\f{|v_{kl}\ind{n}|^2}{r^2}\f{l(l+1)}{r^2}+\f{1}{2}\mu^2\f{|v_{kl}\ind{n}|^2}{r^2} \notag \\
    & \quad\quad\quad\quad\quad\quad +\f{1}{2b^2}\lt(\f{|v_{kl}\ind{n}'|^2}{r^2}-\f{v_{kl}\ind{n}(v_{kl}\ind{n}')^*+v_{kl}\ind{n}'(v_{kl}\ind{n})^*}{r^3}+\f{|v_{kl}\ind{n}|^2}{r^4}\rt)\Bigg) 
\end{align}
by using these two equations \cite{Alcubierre:2018ahf}
\begin{equation}
\begin{aligned}
    \sum_{m=-l}^l Y^{lm*}(\qi,\cphi)Y^{lm}(\qi,\cphi)&=\f{2l+1}{4\pi}, \\
    \sum_{m=-l}^l \n^iY^{lm*}(\qi,\cphi)\n_iY^{lm}(\qi,\cphi)&=\f{1}{r^2}\f{l(l+1)(2l+1)}{4\pi}.
\end{aligned}
\end{equation}
After renormalization, the quantum fluctuation contribution remains finite, and changes the metric non-trivially through the Einstein equation. 

Due to the symmetry of the spacetime, not all components of the Einstein equation are independent. Specifically, we only need to compute two equations in \eref{eq:semiEinstein}
\begin{equation}
\begin{aligned}
    \mbar^2G_0{}^0+\Li&=\langle T_0{}^0\rangle, \\
    \mbar^2(G_0{}^0-G_1{}^1)&=\langle T_0{}^0\rangle-\langle T_1{}^1\rangle,
    \label{eq:metricEq}
\end{aligned}
\end{equation}
where the derivatives $b',s'$ are separated in the above equations, as in \eref{eq:cBSeq}. \eref{eq:KGeq} gives the equation that $v_{kl}\ind{n}$ must satisfy
\begin{equation}
    v_{kl}\ind{n}''+\lt(-\f{2b'}{b}-\f{s'}{s}\rt)v_{kl}\ind{n}'+\lt(\f1r\lt(\f{2b'}{b}+\f{s'}{s}\rt)+b^4s^2(\oi_{kl}\ind{n})^2-b^2\f{l(l+1)}{r^2}-b^2\mu^2\rt)v_{kl}\ind{n}=0.
    \label{eq:scalarEq}
\end{equation}
\eref{eq:scalarEq} is a linear equation with real coefficients, so $v_{kl}\ind{n}$ can be chosen real. In contrast, \eref{eq:metricEq} is highly non-linear. Specific values of the frequencies are required to satisfy these eigenequations. 

To solve \eref{eq:metricEq} and \eref{eq:scalarEq}, suitable boundary conditions are needed. Asymptotic flatness requires 
\begin{equation}
    b(r\to\infty)=1,~~s(r\to\infty)=1,
\end{equation}
and that $v_{kl}\ind{n}/r$ approaches a linear combination of spherical Bessel functions of the first and second kind. Near the origin, if we assume series solutions, we can find that regularity implies $b\approx 1+b_2r^2+\dots$, $s\approx s_0+s_2r^2+\dots$, and $v_{kl}\ind{n}\approx v_{kl,0}\ind{n}r^{l+1}+\dots$, where the coefficients in the expansions are constants. Thus, for \eref{eq:metricEq}, the boundary conditions are 
\begin{equation}
    b(r=0)=1,~~s(r=0)=s_0,
\end{equation}
with $s_0$ a free parameter. The mass $\mu$ and the frequency $\oi$ can still be absorbed, though this is not the case for the frequencies of the mode functions $\oi_{kl}\ind{n}$, which have infinitely many values. 

Because boson stars are localized objects, it is possible to study them by replacing the infinite spatial domain with a finite space region without significantly affecting the solutions. One way to do this is confining the scalar fields within a spherical ball centered at the origin and imposing Dirichlet boundary conditions 
\begin{equation}
    \phi\ind{n}(r=r_M)=0 ~ \Leftrightarrow ~ v_{kl}\ind{n}(r=r_M)=0,
\end{equation}
where $r_M$ is the radius of the ball. If the radius $r_M$ is much larger than the radius of the boson star, the results remain unchanged, as we will show later in \sect{sec:results}. We thus adopt the following boundary conditions for \eref{eq:scalarEq}
\begin{equation}
    v_{kl}\ind{n}(r=0)=v_{kl}\ind{n}(r=r_M)=0.
    \label{eq:quantumBC}
\end{equation}
With these boundary conditions, all the quantum numbers $\{k,l,m\}$ become discrete and we get a real discrete spectrum for $\oi_{kl}\ind{n}$, as guaranteed by the Sturm-Liouville theory \cite{arfken2011mathematical}. 

\subsection{Regularization and renormalization}
\label{sec:PVregularization}

The expectation value of the stress tensor contains quadratic terms of the same operators evaluated at the same point, and is therefore divergent. We use the Pauli-Villars method \cite{Pauli:1949zm} to regularize the divergences. This method incorporates more fields into the Lagrangian \eref{eq:Lagrangian}, where half of them carry an overall minus sign for their contribution to the Lagrangian. This implies that in the stress tensor, their contributions are opposite. When arranging the field masses to be specific values, the divergences can be cancelled. 

These extra fields can be either fictitious fields as a mathematical device to implement regularization, or other physical fields with the desired properties. Despite this, we refer to them collectively as ghost fields in this paper, regardless of their origin. Additionally, their masses should be large enough compared to the physical fields, so that the dynamics of the physical fields is not affected. In a practical sense, their masses should be sent to infinity at the end of the calculation. 

Consider a set of scalar fields $\{\phi^{0r},\phi^{0i},\phi^{1r},\phi^{1i},\dots\}$, with mass $\{m^{0r},m^{0i},m^{1r},m^{1i},\dots\}$ respectively. All of them satisfy the same equation of motion \eref{eq:KGeq}, and can be expanded in the same way as \eref{eq:modeExpansion}. The fields with indices beginning with even numbers contribute oppositely to those with indices beginning with odd numbers. The full stress tensor is the sum of all contributions 
\begin{equation}
    T_{\mn}=\sum_n(-1)^{(n)}T_{\mn}\ind{n},
\end{equation}
where $(n)$ denotes the numerical part of the index $n$. This expression contains quartic, quadratic, and logarithmic divergences; we renormalize the first two dominant contributions, while the last one is neglected because it only has weak dependence on the regulating Pauli-Villars mass. 

The regularization conditions are derived explicitly in FRW spacetime in Appendix~\ref{sec:PVinFRW}. The results are 
\begin{equation}
    \sum_n(-1)^{(n)}=0,~~\sum_n (-1)^{(n)}(m\ind{n})^2=0,~~\sum_n (-1)^{(n)}(m\ind{n})^4=0,
    \label{eq:PVregConditions}
\end{equation}
where the first condition simply implies that half of the fields contribute oppositely. Moreover,  
\begin{equation}
    \langle T_{\mn}\rangle = \langle T_{\mn}\rangle_{\text{fin}} + \di\Li g_{\mn} + \di\mbar^2 G_{\mn},
\end{equation}
where the divergent terms are defined by 
\begin{equation}
\begin{aligned}
    \di\Li&=\f{-1}{32\pi^2}\sum_{n}(-1)^{(n)}(m\ind{n})^4\ln m\ind{n}, \\
    \di\mbar^2&=\f{-1}{48\pi^2}\sum_{n}(-1)^{(n)}(m\ind{n})^2\ln m\ind{n},
    \label{eq:counterterm}
\end{aligned}
\end{equation}
which are identified as the counterterms in $\Li$ and $\mbar^2$ when we consider the renormalization in \eref{eq:counterEinstein}. It is worth mentioning that these results are universal and the same in different spacetimes, because divergences arise from short distance physics, while all spacetimes look Minkowskian locally, as shown in \eref{eq:PVregConditions} and \eref{eq:counterterm} that they only depend on the field masses, not the metric. 

\eref{eq:PVregConditions} admits infinitely many solutions if one is free to incorporate fields arbitrarily. However, we choose to use the minimum number of fields and impose simple relations among their masses. For example, \cite{Berczi:2020nqy,Berczi:2021hdh,Berczi:2024yhb} employ the following set of ghost masses for a real massless physical field $\phi\ind{0}$
\begin{equation}
    m\ind{3}=m\ind{1}=M_{PV},~~m\ind{4}=m\ind{2}=\sqrt{3}M_{PV},~~m\ind{5}=\sqrt{4}M_{PV},
    \label{eq:masslessPV}
\end{equation}
where the Pauli-Villars mass $M_{PV}$ controls the overall magnitude of the ghost fields. Although choosing a different solution is possible, the final results are the same after the masses of the ghost fields are sent to infinity $M_{PV}\to\infty$. 

In our case of two physical fields $\{\phi\ind{0r},\phi\ind{0i}\}$ with the same mass $m\ind{0}=\mu$, we did not find a solution as simple as in \eref{eq:masslessPV}. Nevertheless, a numerical solution exists which, in the limit $m\ind{0}\to0$, is 
\begin{equation}
\begin{aligned}
    & m\ind{1r}=m\ind{1i}=M_{PV},\\
    & m\ind{2r}=m\ind{2i}=\sqrt{\f{11}{2}}M_{PV},\\
    & m\ind{3r}=\sqrt{\f32}M_{PV},m\ind{3i}=\sqrt{\f{15}{2}}M_{PV},
    \label{eq:massivePV}
\end{aligned}
\end{equation}
where the Pauli-Villars mass $M_{PV}$ is supposed to be much larger than the physical field mass $m\ind{0}$. This solution is chosen to use the minimum number of ghost fields and to keep the mass differences among these fields not large. 

To implement the renormalization, we split each bare parameter into a finite renormalized part and a counterterm. The two parameters $\Li=\Li_{\ren}+\di\Li$, $G=G_{\ren}+\di G$ in Einstein's equation \eref{eq:semiEinstein} are sufficient to absorb the quartic and quadratic divergences. The renormalized equation is 
\begin{equation}
    (\mbar^2)_{\ren}G_{\mn}+\Li_{\ren}g_{\mn}=\langle T_{\mn}\rangle-\di\Li g_{\mn}-\di\mbar^2G_{\mn}.
    \label{eq:counterEinstein}
\end{equation}
It follows that if the counterterms are chosen to be the values in \eref{eq:counterterm}, the right-hand side becomes the finite part of the stress tensor, which can be interpreted as the renormalized stress tensor
\begin{equation}
    \langle T_{\mn}\rangle_{\ren} = \langle T_{\mn}\rangle_{\text{fin}}.
\end{equation}
Hence, 
\begin{equation}
    (\mbar^2)_{\ren}G_{\mn}+\Li_{\ren}g_{\mn}=\langle T_{\mn}\rangle_{\ren},
    \label{eq:renEinstein}
\end{equation}
where all these renormalized quantities correspond to physical, measurable values\footnote{A rigorous derivation of the renormalized stress tensor has been obtained in algebraic quantum field theory \cite{Hollands:2001nf,Hollands:2001fb,Hollands:2002ux,Hollands:2004yh,Hollands:2009bke}.}.
For static and asymptotically flat spacetimes, it is reasonable to set $\Li_{\ren}=0$.

If the higher-order curvature terms are included, additional terms appear in the semiclassical Einstein equation, which can be collectively written as $a H_{\mn}$, where $H_{\mn}$ is a purely geometric tensor. The renormalization scale, i.e. $M_{PV}$, is chosen such that the renormalized parameter vanishes $a_\ren=0$. Choosing a different scale would shift this relation, but $\di a$ is only logarithmically dependent on $M_{PV}$, which is subdominant compared to those in \eref{eq:counterterm} and therefore can be safely neglected \cite{Tranberg:2008ae}. Within this setup, the renormalized Einstein equation reduces to \eref{eq:renEinstein}. 

The renormalization only affects the divergent part in \eref{eq:stressCQ}, so we can define the renormalized quantum fluctuation $(T_{\mn}^{(\text{q})})_{\ren}$
\begin{equation}
    \langle T_{\mn}\rangle_{\ren} = T_{\mn}^{(\text{c})}+(T_{\mn}^{(\text{q})})_{\ren}.
\end{equation}
The renormalized quantum energy density and radial pressure can be defined as  
\begin{equation}
    (\ri^{(\text{q})})_{\ren}=-(T_0{}^0{}^{(\text{q})})_{\ren},~~
    (p_r^{(\text{q})})_{\ren}=(T_1{}^1{}^{(\text{q})})_{\ren},
    \label{eq:renDensity}
\end{equation}
which are the main objects to be computed in this paper, and the associated counterterms are 
\begin{equation}
\begin{aligned}
    \di\ri  = -\di\Li-\di\mbar^2 G_0{}^0,~~
    \di p_r  = \di\Li+\di\mbar^2 G_1{}^1.
\end{aligned}
\end{equation}

The ghost fields satisfy the same equation of motion as the physical fields, though they have different masses. In particular, the quantum fluctuations of their stress tensors are static and spherically symmetric. Because the physical fields already account for the entire classical part in \eref{eq:stressCQ}, we set the expectation values of the ghost fields in the coherent state to be 
\begin{equation}
    \langle\chi|\phi\ind{n}|\chi\rangle=0,~~(n)>0,
\end{equation}
which implies that these heavy fields are not excited and contribute only the quantum fluctuations; their job is solely to remove the ultraviolet divergences.

Among the various available regularization schemes, the Pauli-Villars method has several advantages. First, it preserves diffeomorphism invariance explicitly, because the modification is applied at the level of the Lagrangian. This is not the case for some methods, for example normal ordering, though simple, is not a good way to explore curved spacetime effects. Second, it does the regularization mode by mode, which is highly suitable for numerical calculation when analytic techniques are not feasible. Third, it can be readily used in generic spacetimes, even if they are dynamical. Finally, it is worth emphasizing that different regularization schemes could also be used, and equivalent physical results should be obtained.

\section{Quantum field and stress tensor}
\label{sec:results}

In this section, we first discuss the numerical method we use to solve for the classical boson star metric and the quantum mode functions, whose results are reported later. The stress tensor is obtained for a large range of solutions. All quantities are dimensionless by multiplying appropriate powers of the physical field mass $\mu$ like in \eref{eq:dimensionless}, effectively $\mu=1$, and we omit the $\sim$ for simplicity. Readers who are not interested in the numerical details can skip \sect{sec:code}. 

\subsection{Numerical setup}
\label{sec:code}

To compute the stress tensor in a boson star spacetime, we need to first solve \eref{eq:cBSeq} for the metric functions $b,s$, which form a non-linear system of equations, and then solve \eref{eq:scalarEq} in the resulting background, which is a linear eigenequation. Both problems are solved numerically, where derivatives are computed using the spectral method \cite{trefethen2000spectral,boyd2001chebyshev}, in which fields are represented using a set of orthogonal smooth basis functions on unevenly spaced grids, Chebyshev grids. The differential equations then become matrix equations, analogous to using the finite difference method. The most significant advantage of the spectral method is that numerical errors decrease exponentially as the number of grid points increases, unlike the finite difference method where errors decrease only as a power law. Details of implementing the spectral method are provided in Appendix~\ref{sec:spectral}. 

Given boundary conditions, the non-linear \eref{eq:cBSeq} is solved with the Newton-Raphson method \cite{press2007numerical}, and the linear \eref{eq:scalarEq} is solved by evaluating eigenvalues and eigenfunctions of the matrices using standard techniques. We start with the discussion of the linear eigenvalue problem, which highlights more technical details. 

\begin{figure}[tbp]
    \centering
    \includegraphics[width=0.43\linewidth]{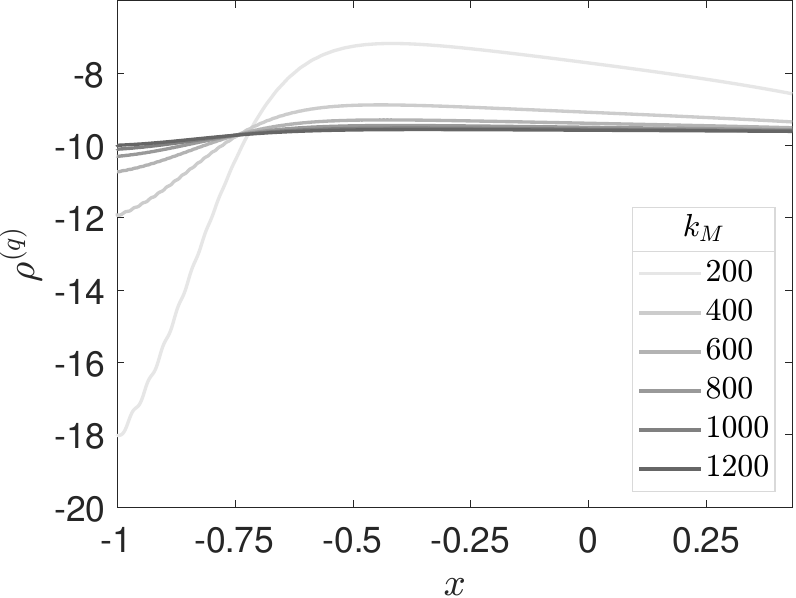}
    \hfill
    \includegraphics[width=0.44\linewidth]{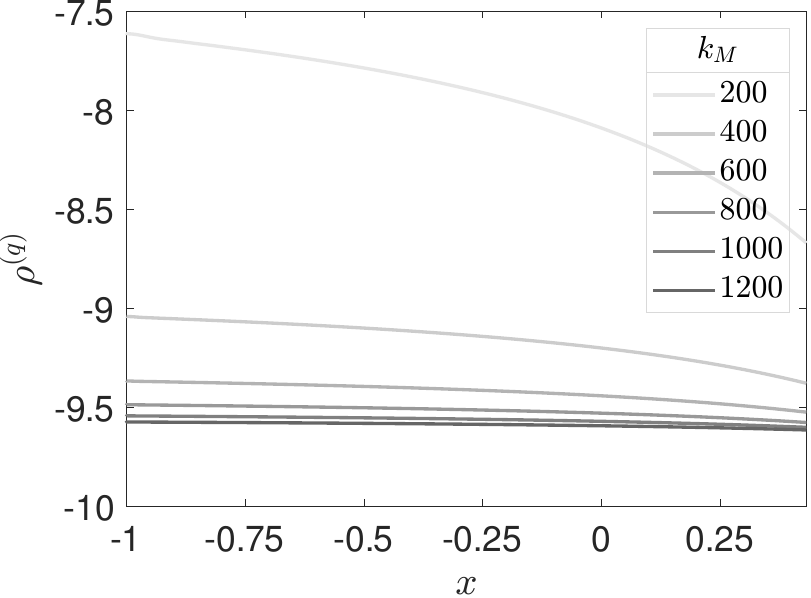}
    \includegraphics[width=0.46\linewidth]{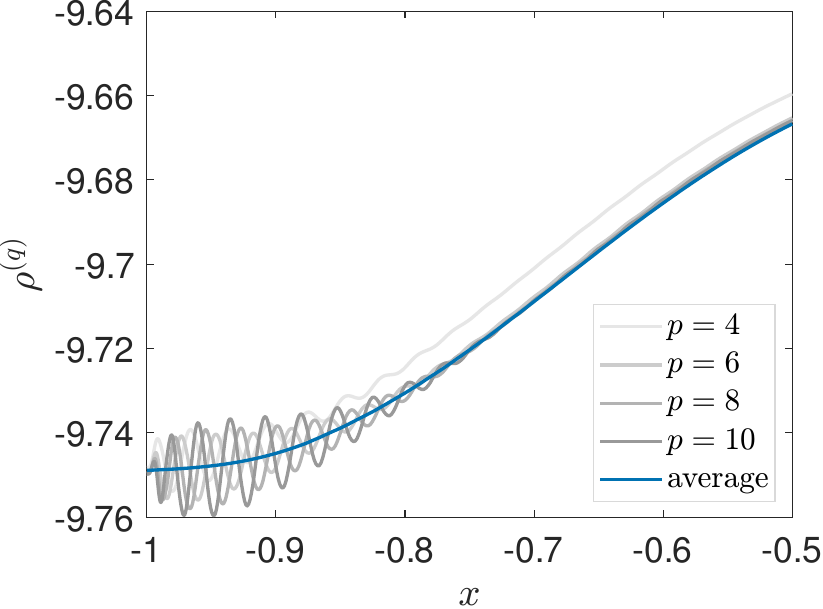}
    \hfill
    \includegraphics[width=0.45\linewidth]{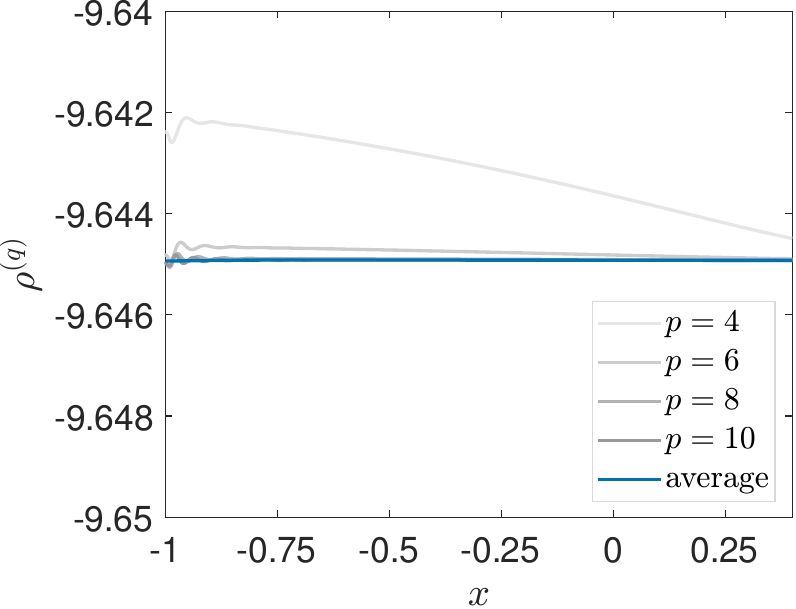}
    \caption{Quantum energy density from summation over different numbers of modes (upper) and extrapolation to the infinite-mode limit using polynomials of different degrees (lower) for the maximum compactness boson star spacetime (left) and Minkowski spacetime (right).}
    \label{fig:extrapolation}
\end{figure}

As discussed in \sect{sec:sphere}, the boundary conditions we use are \eref{eq:quantumBC}, which confine the scalar fields within a finite spherical region. The eigenvalues of \eref{eq:scalarEq} are therefore discrete, and the eigenfunctions can be distinguished by their number of zeros. The finite-size boundary confining the scalar fields does not affect the physical results, because the spacetime approaches Minkowski spacetime rapidly at large $r$, and thus the stress tensor outside $r_M$ is well-controlled. Moreover, the modification of the boundary to the stress tensor occurs only near the spherical surface and decays rapidly away from it \cite{Birrell:1982ix}. We verify this explicitly later in Minkowski spacetime (see \fig{fig:extrapolation}). 

For eigenfunctions with a large number of zeros, equivalent to large eigenvalues $\oi_{kl}^n$, the functions oscillate rapidly along the radial direction. To implement the spectral method, the continuous interval should be replaced by Chebyshev grids. Due to these rapid oscillations, it is important both to use sufficiently many grid points and to select a sphere of adequate size, so that the eigenfunctions can be represented well on the grid. The radius $r_M$ of the sphere should be much larger than the size of the boson star, which we choose to be $r_M=32$; we have checked that the results do not change significantly when varying the radius to larger values. The number of grid points determines how many modes can be included in the stress tensor calculation, and we use $3072$ points so that more than $1200$ modes with different $k$ for each $l$ can be used to compute the stress tensor. 

Unlike the scalar fields, the metric function $b$ decays as a power law (see \fig{fig:cBSpoints}), so a large interval is needed to represent it accurately. We transform the infinite interval $r\in[0,\infty)$ to a finite compact interval $x\in[-1,1]$ by compressing the large $r$ region using this coordinate transformation
\begin{equation}
    r = \f{r_0(1+x)}{1-x+2r_0/r_B}.
    \label{eq:r2x}
\end{equation}
When $x=1$, $r=r_B$, which is a finite number to avoid the singularity at infinity, but is larger than the sphere confining the scalar fields. The numerical value we took was $r_B=2048$ so that the error caused by this boundary is negligible and effectively what we compactify is the infinite interval. $r_0$ controls how the interval is compressed and a large $r_0$ implies that the small $r$ region is compressed more; the numerical value we chose was $r_0=4$, so the boundary of the scalar fields is located at $x=0.78$. 

With this finite interval and the boundary conditions in \eref{eq:classicalBC}, we solve \eref{eq:cBSeq} for classical boson stars via the Newton-Raphson method. The resulting metric is then used to solve for mode functions, as we showed previously. The Chebyshev grids are used again, but in different coordinates, thus any quantities computed in the $r$ coordinate should be transformed accordingly. In particular, the stress tensor should be transformed. We will report most of the results in the $x$ coordinate, which magnifies the interior region of the boson stars where the relevant physics occurs. 

As discussed in \sect{sec:PVregularization}, the Pauli-Villars ghost fields are required to eliminate divergences, and their masses should be sufficiently large. In our numerical setup $M_{PV}=5$, meaning that the lightest ghost field is five times heavier than the physical fields. Larger values of $M_{PV}$ are possible, but more modes are required to compute the stress tensor due to the increased mass hierarchy. We checked the results with $M_{PV}=2,3,4,5,6$, and all are qualitatively identical. For consistency, we report results with $M_{PV}=5$ throughout this paper. 

\begin{figure}[tbp]
    \centering
    \includegraphics[width=0.45\linewidth]{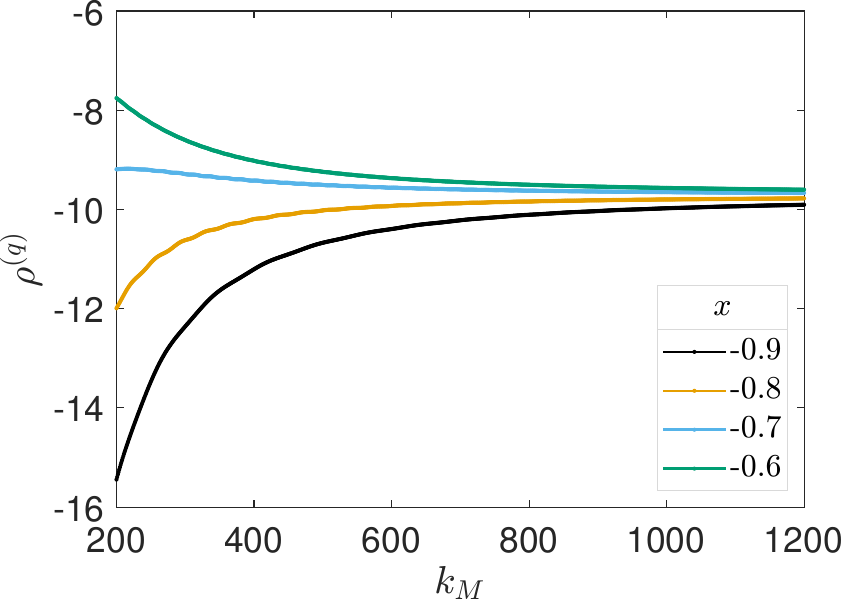}
    \hfill
    \includegraphics[width=0.45\linewidth]{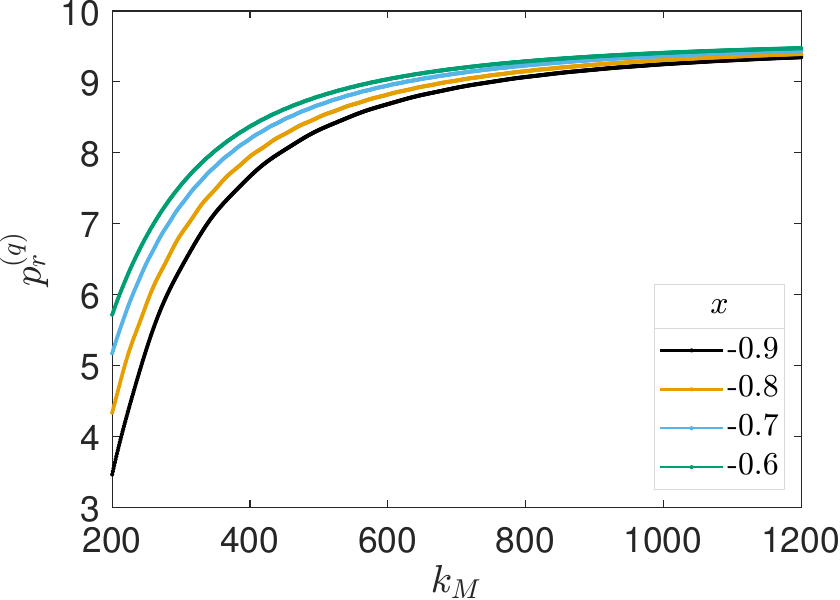}
    \caption{Quantum energy density (left) and radial pressure (right) at fixed locations, obtained from summation over different numbers of modes, for the maximum compactness boson star spacetime.}
    \label{fig:convergence}
\end{figure}

A further numerical consideration is that only finitely many modes can be included. Although the Pauli-Villars method naturally preserves the diffeomorphism invariance, truncating the momentum space sum breaks it. To restore this symmetry, we compute the stress tensor with different numbers of modes, say
\begin{equation}
    T_{\mn}^{\text{(q)}}(x;k_M,l_M)=\sum_{k=1}^{k_M}\sum_{l=0}^{l_M}\sum_{m=-l}^{l}\sum_n(-1)^{(n)}\mcl{T}_{\mn}\ind{n}(f_{klm},f_{klm}^*)
\end{equation}
adds all modes up to $k=k_M$ and $l=l_M$. The result approaches the infinite-mode limit closer and closer as more and more modes are involved, which provides an appropriate basis for extrapolation. This process can be thought of as a series expansion about $(1/k_M=0,1/l_M=0)$, and diffeomorphism is restored when this limit is reached. 

Although $(1/k_M,1/l_M)$ spans a two-dimensional parameter space, the problem can be simplified by restricting to a single trajectory in this space, and a one-dimensional extrapolation is more accurate and simpler to implement. Specifically, we fix the ratio $l_M/k_M$ to be a fixed number, say $2$, and the infinite-mode limit becomes $1/k_M\to0$. Some representative results are shown in the upper two panels of \fig{fig:extrapolation} and in \fig{fig:convergence}. From the latter, one can clearly see that the stress tensor at different spatial points converges rapidly as $k_M$ increases, while the former illustrates the overall convergence in the central regions of the spacetimes. 

The extrapolation is implemented by fitting the large momentum part of the stress tensor data with polynomials in $1/k_M$ using the least squares method. Only even power terms appear because the stress tensor only contains quadratic terms of the momentum. In other words, 
\begin{equation}
    T_{\mn}^{\text{(q)}}(x;k_M,2k_M)\simeq a_{\mn,0}(x)+\f{a_{\mn,2}(x)}{k_M^2}+\f{a_{\mn,4}(x)}{k_M^4}+\cdots+\f{a_{\mn,p}(x)}{k_M^p}
\end{equation}
for large $k_M$, and the coefficient 
\begin{equation}
    a_{\mn,0}(x)=\lim_{k_M\to\infty}T_{\mn}^{\text{(q)}}(x;k_M,2k_M)
\end{equation}
is the limiting value. In the lower two panels of \fig{fig:extrapolation}, the data from $k_M=200$ to $k_M=1200$, including one thousand $T_{\mn}^{\text{(q)}}(x;k_M,2k_M)$, is extrapolated using polynomials of different degrees. Numerically, we find that polynomials with different highest order $p$ yield the same results (upon neglecting the oscillations, which can be removed by the averaging method described below), if $p$ is within a sequence of large consecutive integers; we use eighth-order polynomials in practice. 

Different $l_M/k_M$ ratios produce the same extrapolated results, which we have checked numerically; since the function is sufficiently smooth near the infinite-mode limit, the results should be path-insensitive. Due to numerical errors, the extrapolation curves show many oscillations, which can be treated as error bars for our data. Although including more modes can reduce these oscillations, it is not likely to eliminate them completely. To mitigate these numerical errors, we apply a final averaging to the extrapolation data, which uses a Gaussian-weighted moving average method; a sample window covering a range of ten oscillations is good enough to remove these oscillations. The final results are shown in the lower two panels in \fig{fig:extrapolation} for the maximum compactness boson star metric and the Minkowski metric, where for the former the energy density is a smooth function of $x$ and for the latter it is a constant, which can be completely cancelled by the cosmological constant in renormalization. 

\subsection{Boson star metric}
\label{sec:metric}

\begin{figure}[tbp]
    \centering
    \includegraphics[width=1\linewidth]{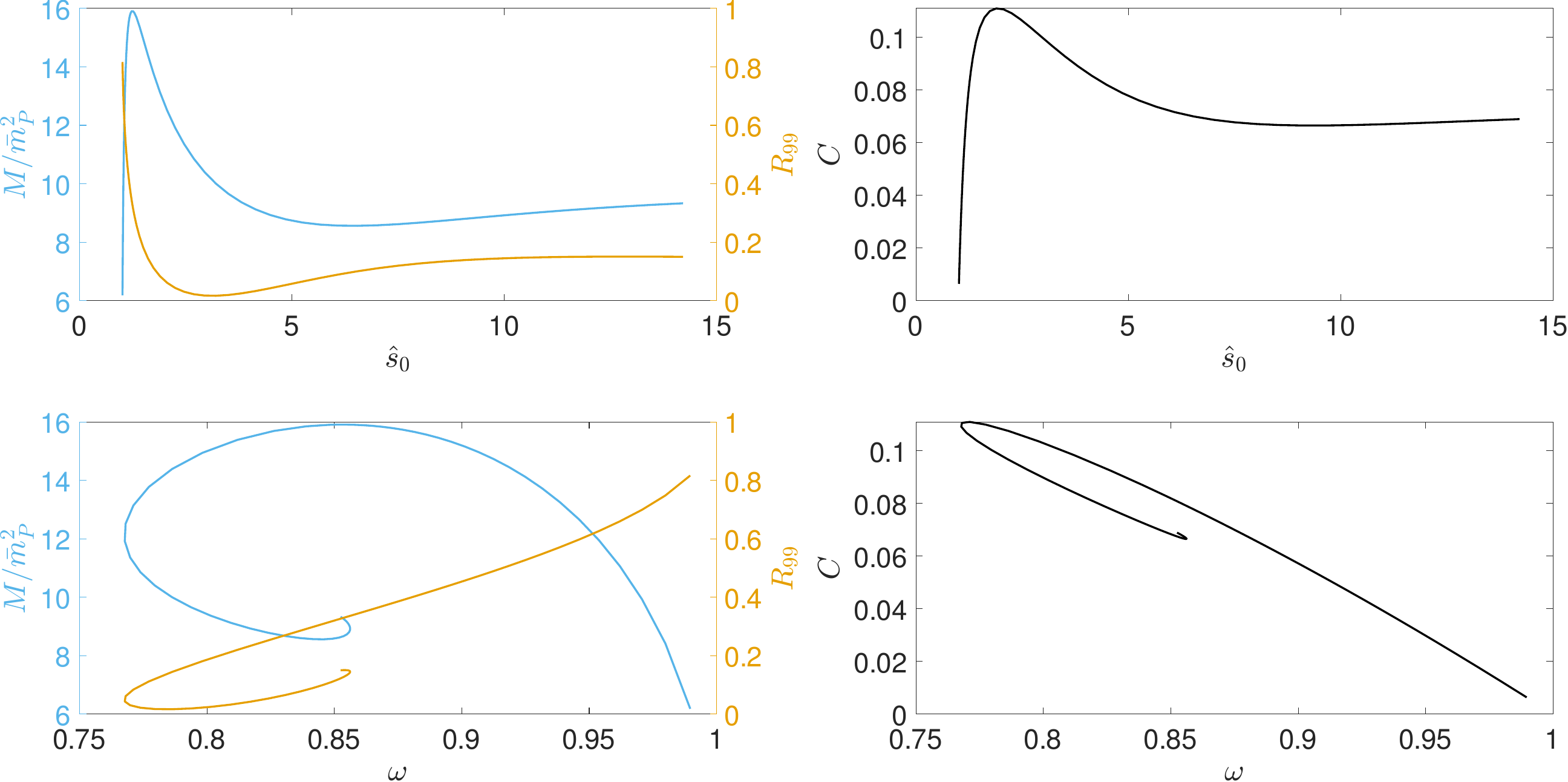}
    \caption{Mass, radius, and compactness of classical boson stars. The upper two panels use the monotonic parameter $\hat{s}_0$ as their horizontal axes, while the lower two panels use the frequency $\omega$.}
    \label{fig:cBSglobal}
\end{figure}

The boson star metric is obtained by solving \eref{eq:cBSeq} with $\Li=0$. We assume $\si$ to be a monotonically decreasing function of $r$. As discussed in \sect{sec:classicalBS}, this system contains only one free parameter which can be chosen as $s_0$, $\si_0$, $\oi$ and so on. However, not all of these choices provide a single-valued mapping to the solutions; for example, multiple boson star solutions can correspond to the same frequency $\oi$. In this paper, we adopt the value at the origin $\h{s}_0=s_0\oi$ defined in \eref{eq:hats} as the unique free parameter, which establishes a monotonic correspondence with the solutions (see \fig{fig:cBSglobal}). 

\begin{figure}[tbp]
    \centering
    \includegraphics[width=1\linewidth]{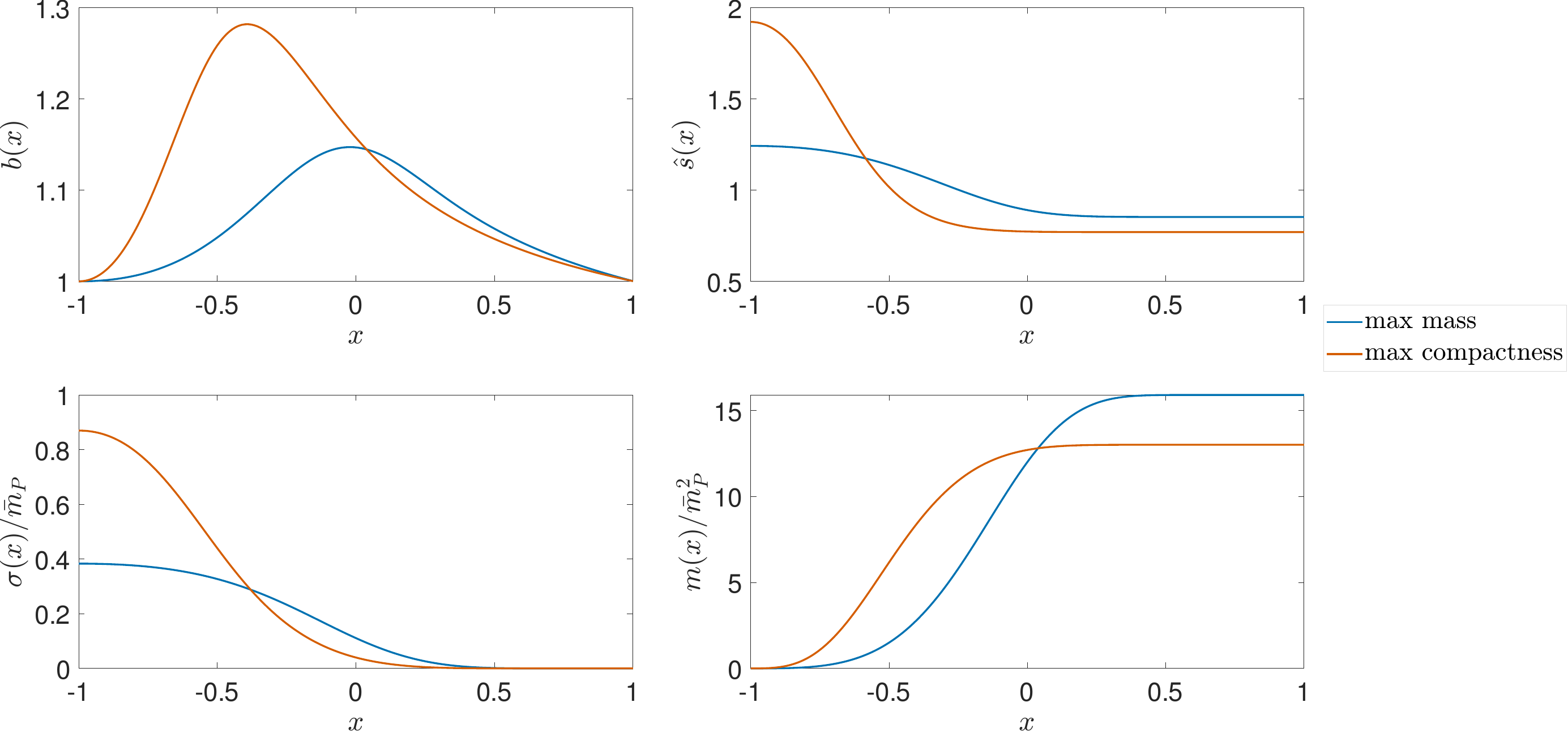}
    \caption{Boson star solutions at maximum mass ($\hat{s}_0=1.24$) and maximum compactness ($\hat{s}_0=1.92$).}
    \label{fig:cBSpoints}
\end{figure}

\fig{fig:cBSpoints} presents two representative solutions in the compact coordinate \eref{eq:r2x}. The blue curve corresponds to the maximum mass solution, while the red curve corresponds to the maximum compactness solution. The metric function $b$ approaches unity at both $x=-1$ and $x=1$, consistent with the condition of regularity at the origin and asymptotic flatness at infinity, and becomes larger than unity within the interior region. $\h{s}$ is a decreasing function of $x$ and approaches the frequency $\oi$ rapidly, implying that $s$ approaches unity rapidly, as required for asymptotic flatness. The lower left panel displays the localized scalar field profile, and the final panel shows the monotonically increasing mass function $m(x)$, which attains the total mass as the scalar field vanishes. 

A comparison of the two solutions shows that all the fields of the maximum compactness solution are concentrated more tightly near the origin. The metric and the scalar field exhibit greater variations in their values. Although the mass function of this solution finally reaches a smaller value than that of the maximum mass solution, its radius is significantly reduced, resulting in a larger compactness. 

\fig{fig:cBSglobal} illustrates how the mass, radius and compactness vary across the family of solutions. In the upper two panels, the horizontal axis is $\h{s}_0$. One finds that $\h{s}_0$ is a monotonic parameter and can effectively denote different solutions. The mass (light blue curve) increases rapidly as $\h{s}_0$ deviates from unity, which corresponds to the Minkowski limit; the maximum mass solution in \fig{fig:cBSpoints} occurs at $\h{s}_0=1.24$. Beyond this point, the mass decreases until it reaches a minimum point, and it finally varies slowly around a constant value. 

The radius (orange curve) exhibits a similar but opposite trend. It decreases initially and increases after a point on the right of the maximum mass point. It finally varies slowly around a constant value. The compactness follows a pattern very similar to the mass, but the maximum compactness solution in \fig{fig:cBSpoints} is reached later than the maximum mass point, at $\h{s}_0=1.92$. This is clearly from the mismatch of the mass and the radius, reaching extreme points at different $\h{s}_0$, which ultimately stems from the non-linearity of the system. 

The lower two panels show the same quantities, but with the horizontal axis replaced by the frequency $\oi$. The mass increases from $0$ near the lower right corner, and forms a contracting spiral approaching a fixed point in the center. The radius also traces a spiral but it decreases initially, while the compactness exhibits a noticeably eccentric spiral. These imply that $\oi$ is not a monotonic parameter as $\h{s}_0$ and solutions with different mass can possess the same $\oi$. However, we include these two panels to facilitate comparison with earlier work. 

\begin{figure}[tbp]
    \centering
    \includegraphics[width=0.5\linewidth]{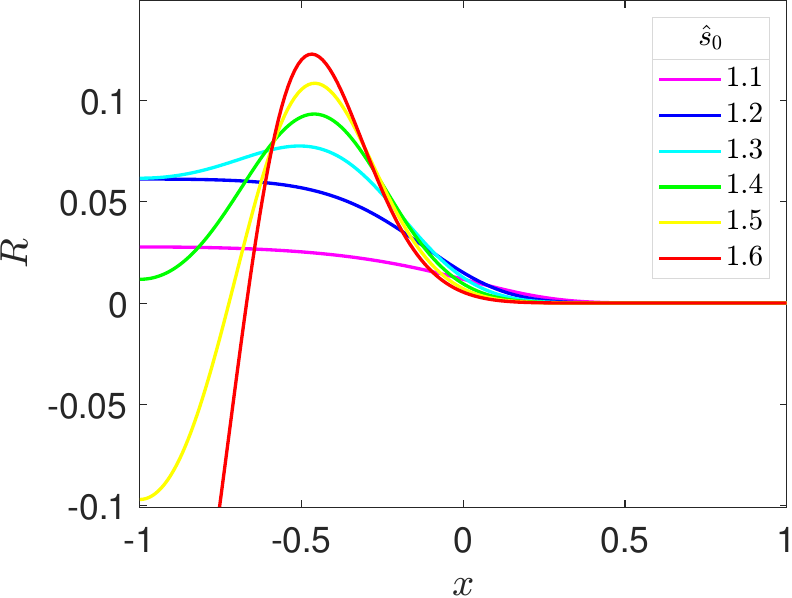}
    \caption{Ricci scalar of classical boson stars.}
    \label{fig:ricci}
\end{figure}

In addition to the mass and compactness, the Ricci scalar $R$ is another important characteristic of boson star geometries, which quantifies the degree of spacetime curvature. \fig{fig:ricci} shows the Ricci scalar for several solutions with different $\h{s}_0$. For small $\h{s}_0$, $R$ is localized near the center and its magnitude is small. As $\h{s}_0$ increases, $R$ becomes larger but it is no longer a monotonic function; the maximum point shifts to $x=-0.5$. For larger values of $\h{s}_0$, $R$ near the center becomes increasingly negative, and the maximum absolute value of $R$ is again attained in the center. 

This behaviour continues for larger $\h{s}_0$, and the maximum absolute value located in the center keeps increasing without apparent bound and becomes much larger than its value elsewhere. This is in contrast with other quantities such as the mass and compactness, which remain bounded and undergo alternating increases and decreases. Later in \sect{sec:stressTensor}, we will show that the curvature is directly related to the magnitude of quantum effects and thus serves as a reliable measure for them. 

\subsection{Mode functions}
\label{sec:modeFunction}

The mode functions are obtained by solving the linear eigenequation \eref{eq:scalarEq} with the metric held fixed. Due to the boundary conditions \eref{eq:quantumBC}, the eigenfrequencies are positive and discrete, and the eigenfunction can be chosen to be real. The eigenfunctions can be further listed according to the number of zeros within the interval; the first one has no zeros, and the subsequent ones exhibit an consecutively increasing number of zeros, analogous to the case of hydrogen atom. 

\begin{figure}[tbp]
    \centering
    \includegraphics[width=0.48\linewidth]{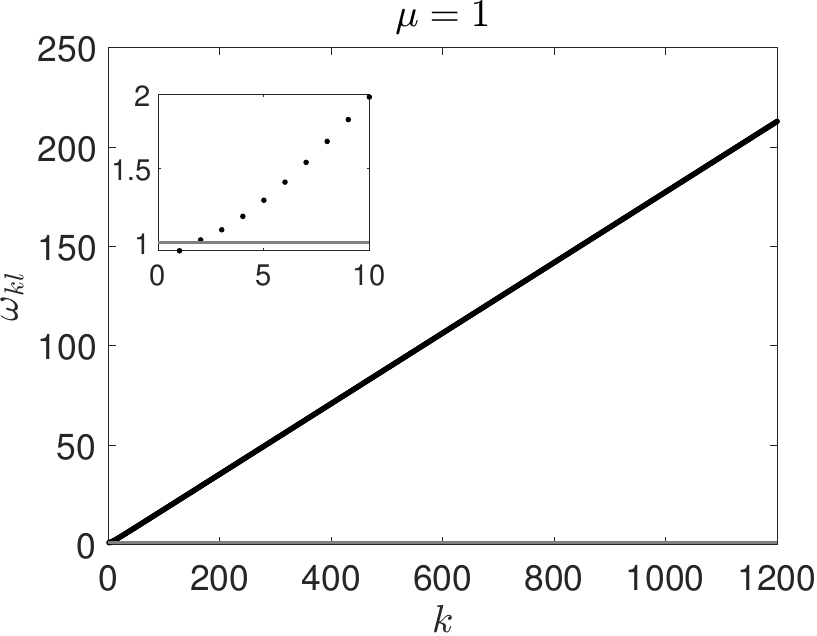}
    \hfill
    \includegraphics[width=0.48\linewidth]{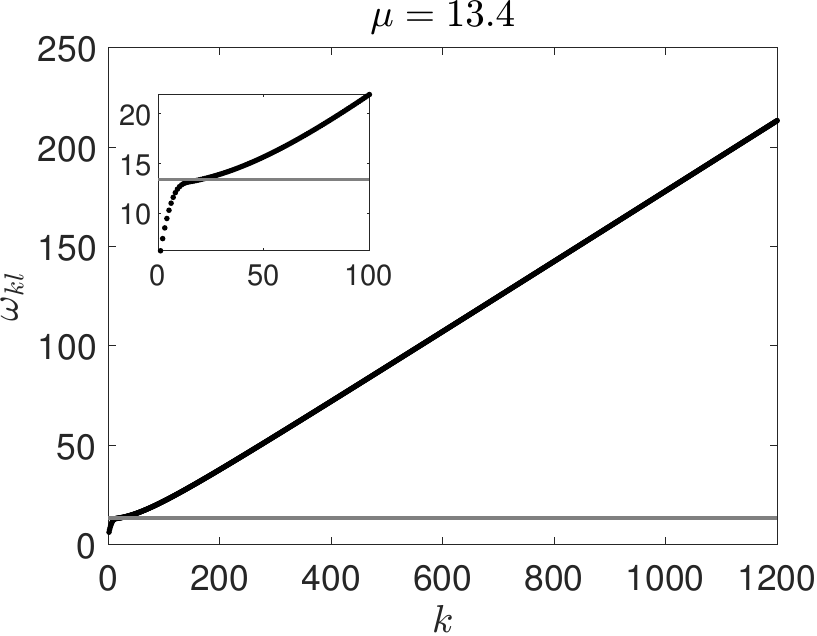}
    \caption{Frequency spectra for the physical field (left) and the most massive ghost field (right) in the metric with maximum compactness.}
    \label{fig:spectrum}
\end{figure}

\fig{fig:spectrum} displays the frequency spectra for two representative fields with different masses, one light and the other one very heavy, which correspond to the physical field and the most massive ghost field respectively. Although the spectrum depends on two quantum numbers $k$ and $l$, we fix the ratio $l/k=2$ for illustration, because increasing either of them leads to high momentum modes. The qualitative behaviours are similar in both cases. The frequency increases monotonically with $k$ and quickly exceeds the mass threshold (gray line). In the large momentum regime, the frequency follows a straight line, which is not sensitive to the field mass as the momentum already dominates. 

\begin{figure}[tbp]
    \centering
    \includegraphics[width=1\linewidth]{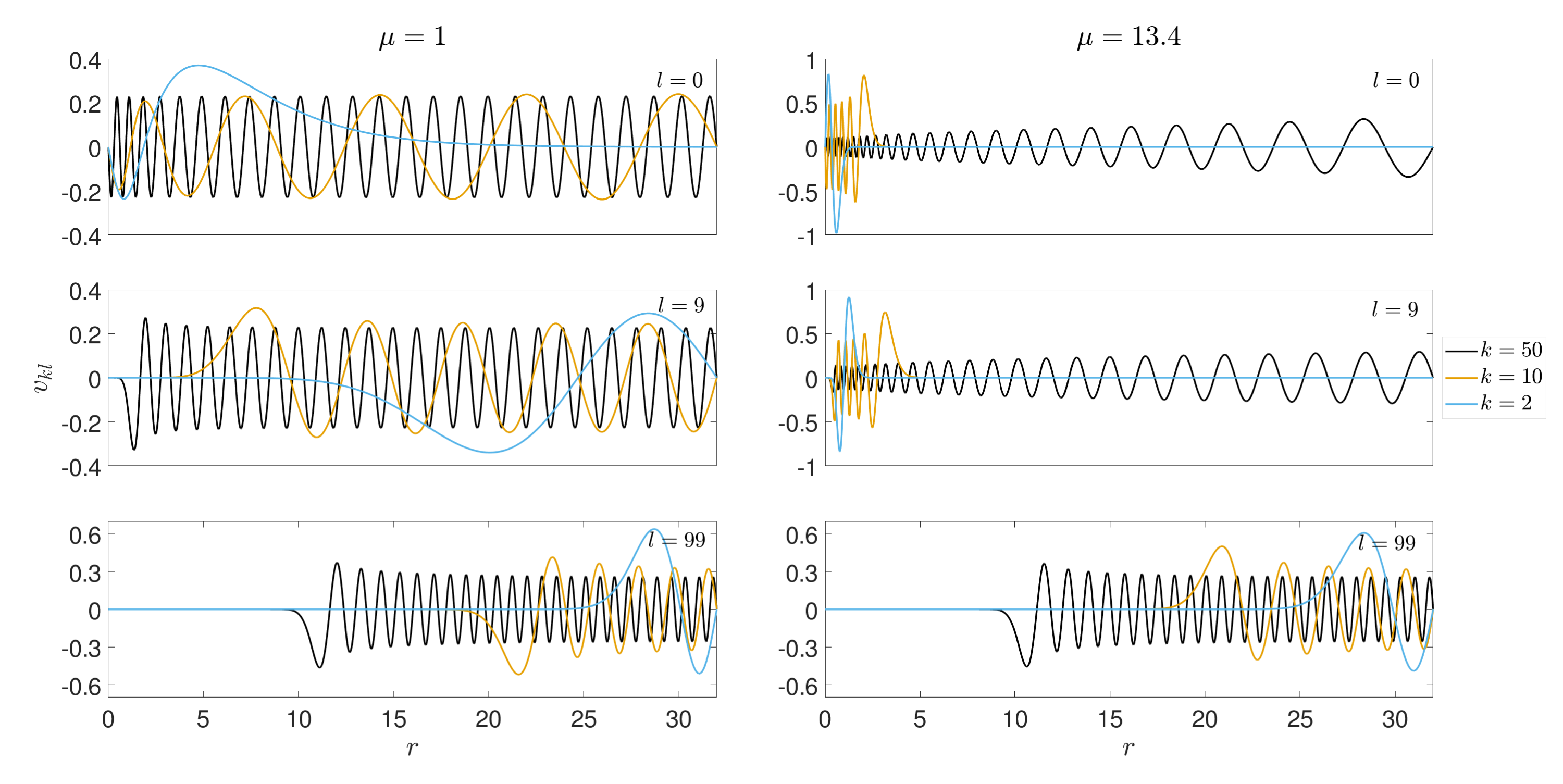}
    \caption{Mode functions $v_{kl}$ with various $k$ and $l$ of the physical field (left) and the most massive ghost field (right) in the metric with maximum compactness.}
    \label{fig:modeFunctions}
\end{figure}

The insets show enlarged views of the first few frequencies of the spectra. For the massive field, several frequencies lie below the mass threshold, corresponding to bound modes that decay exponentially away from the center. Modes with frequencies exceeding the mass are scattering modes and their derivatives at the outer boundary are not zero, though they satisfy the Dirichlet boundary conditions there (see \fig{fig:modeFunctions}). Owing to the finite interval, the mass does not provide a sharp separation between bound modes and scattering modes, and some modes slightly below the mass behave as scattering states, as indicated by their oscillatory behaviour near the boundary. However, this distinction becomes negligible in the infinite-volume limit and does not affect the numerical results presented here. As expected, the massive field supports a significantly larger number of bound states than the light field. 

Representative mode functions $v_{kl}$ with various $k$ and $l$ are shown in \fig{fig:modeFunctions}; to illustrate the behaviour of scattering modes more clearly, we present the results in the $r$ coordinate. Taking the first panel as an example, where $l=0$, it shows that the $k=2$ mode is bound while the $k=10$ and $k=50$ modes are both scattering, and exhibit trigonometric oscillation patterns at larger $r$. As $l$ increases, the functions develop an increasingly large region near the center where they vanish. For the massive field, the bound modes decay more rapidly at larger $r$. In the large $l$ regime, say $l=99$, the mode functions of the different fields become nearly indistinguishable (note that adding an overall minus sign does not change the mode functions because they enter the stress tensor quadratically). 

Although the results in this subsection are all obtained using the metric with maximum compactness, other boson star metrics produce qualitatively similar spectra and mode function behaviour.  

\subsection{Stress tensor}
\label{sec:stressTensor}

As discussed in \sect{sec:metric}, the only parameter in this system that changes the physical properties of the solutions is $\h{s}_0$, we therefore focus on how the stress tensor varies as $\h{s}_0$ is changed. 

\begin{figure}[tbp]
    \centering
    \includegraphics[width=0.48\linewidth]{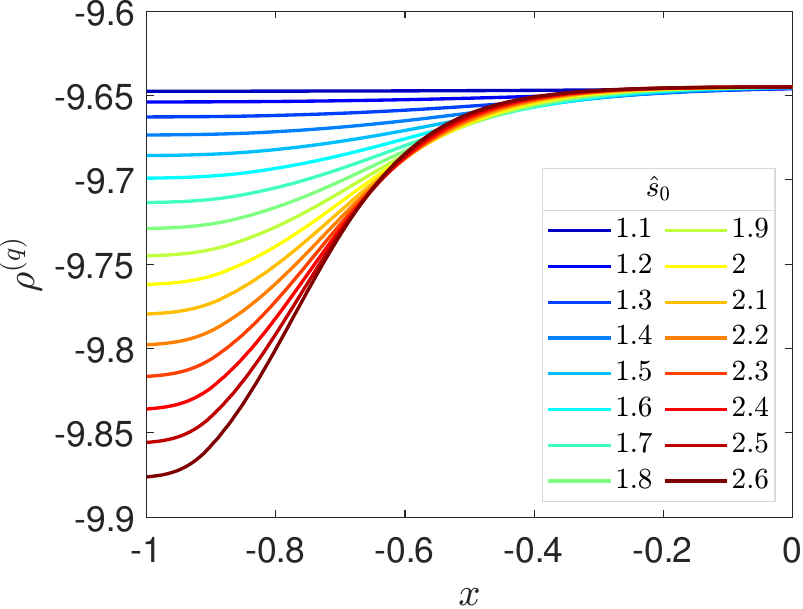}
    \hfill
    \includegraphics[width=0.48\linewidth]{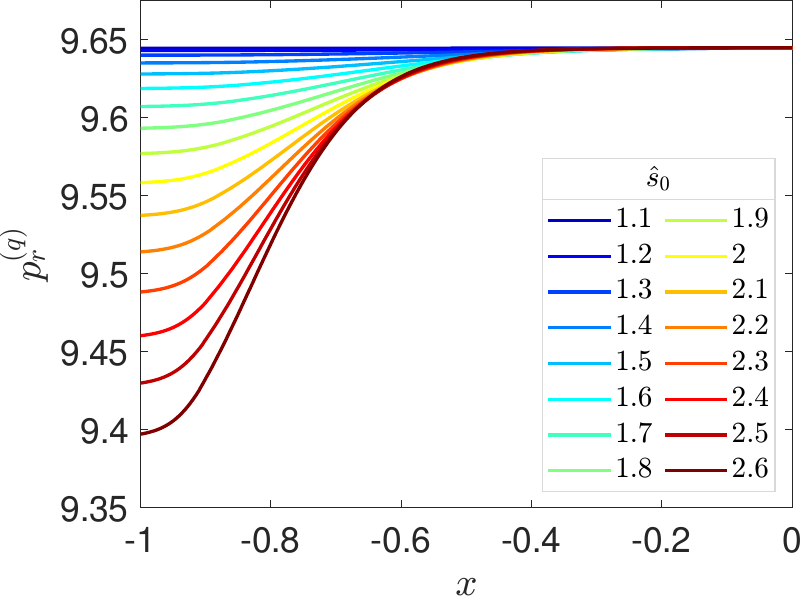}
    \caption{Bare quantum energy density and radial pressure.}
    \label{fig:bareStress}
\end{figure}

\fig{fig:bareStress} shows the bare quantum energy density and radial pressure defined in \eref{eq:quantumDensity} for a set of solutions with different $\h{s}_0$, obtained by extrapolating to the infinite-mode limit. The left panel shows the energy density, whose bare value is negative. When the boson star spacetime is close to Minkowski spacetime (small $\h{s}_0$), the energy density is nearly constant as a function of $x$. As $\h{s}_0$ increases, the minimum value in the center becomes increasingly negative, reflecting the growing influence caused by the curved spacetime. However, far from the boson star, the spacetimes become asymptotically flat, and the quantum effects become negligible for all solutions. 

The behaviour of the radial pressure is similar, except that its bare value is positive. The value in the center becomes smaller as $\h{s}_0$ increases. Far from the center, the spacetimes become flat and the quantum effects vanish. The energy density and pressure approach a value that is equal in magnitude ($\sim9.65$), but opposite in sign, indicating that they can be cancelled by a single cosmological constant, and the resultant Minkowski vacuum energy and pressure vanish, as expected. 

\begin{figure}[tbp]
    \centering
    \includegraphics[width=0.48\linewidth]{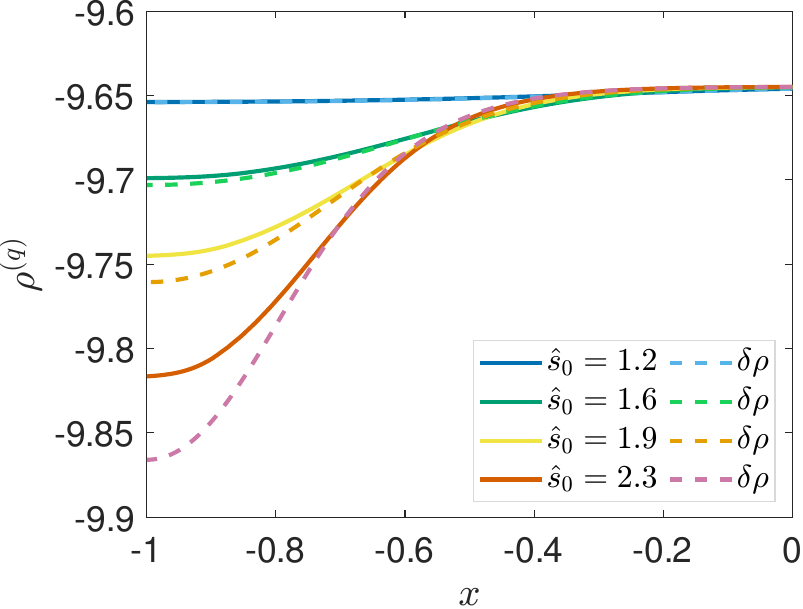}
    \hfill
    \includegraphics[width=0.48\linewidth]{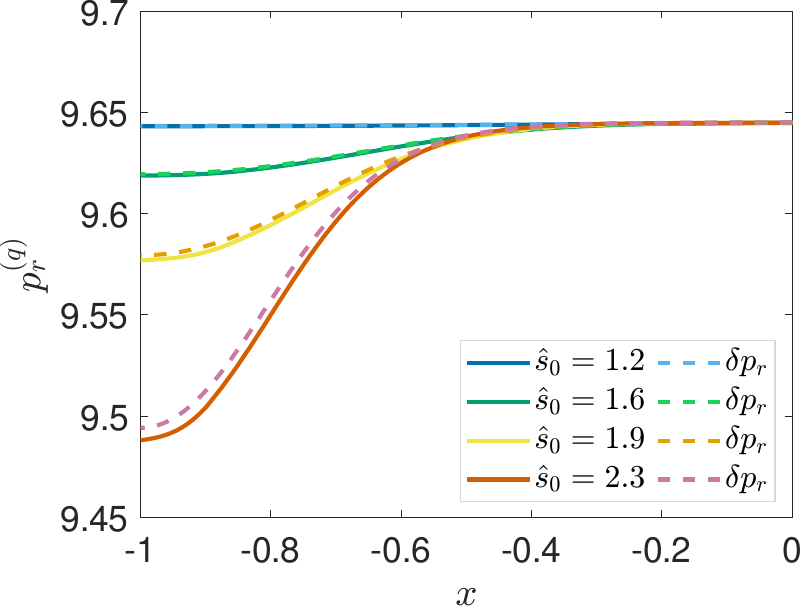}
    \caption{Bare quantum energy density, radial pressure (solid lines) and their counterterms (dashed lines with similar colors).}
    \label{fig:counterStress}
\end{figure}

In the renormalization procedure, the counterterm is subtracted from the bare stress tensor as in \eref{eq:counterEinstein}. \fig{fig:counterStress} compares these two quantities for several representative solutions; each bare tensor is shown with solid lines, while the counterterm associated with this bare tensor is plotted in dashed lines of similar color. The counterterm is different from the bare tensor near the origin, whereas they are the same for large $x$. Also, it can be seen that for small $\h{s}_0$, the solid line and the dashed line nearly coincide, implying that the renormalized tensor is approximately zero. Their discrepancy in the center grows monotonically as $\h{s}_0$ increases, leading to a larger renormalized quantum stress tensor. 

It is instructive to consider the reason for this enhancement. Although one might expect the mass to play this role, the solution with the largest mass $\h{s}_0=1.2$ exhibits the most indistinguishable quantum effect among the four examples. A similar conclusion applies to the case of compactness, where the $\h{s}_0=1.9$ solution has the largest compactness, but the quantum effect of $\h{s}_0=2.3$ is clearly stronger. A more appropriate quantity is the Ricci scalar $R$, as shown in \fig{fig:ricci} that its maximum absolute value is a monotonically increasing quantity of $\h{s}_0$, indicating that larger curvature induces a larger quantum fluctuation in stress tensor. 

\begin{figure}[tbp]
    \centering
    \includegraphics[width=0.48\linewidth]{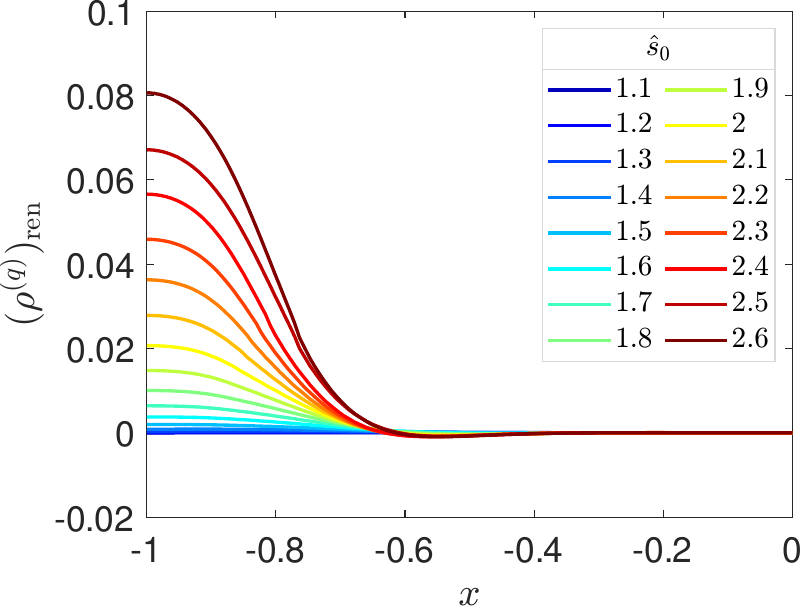}
    \hfill
    \includegraphics[width=0.48\linewidth]{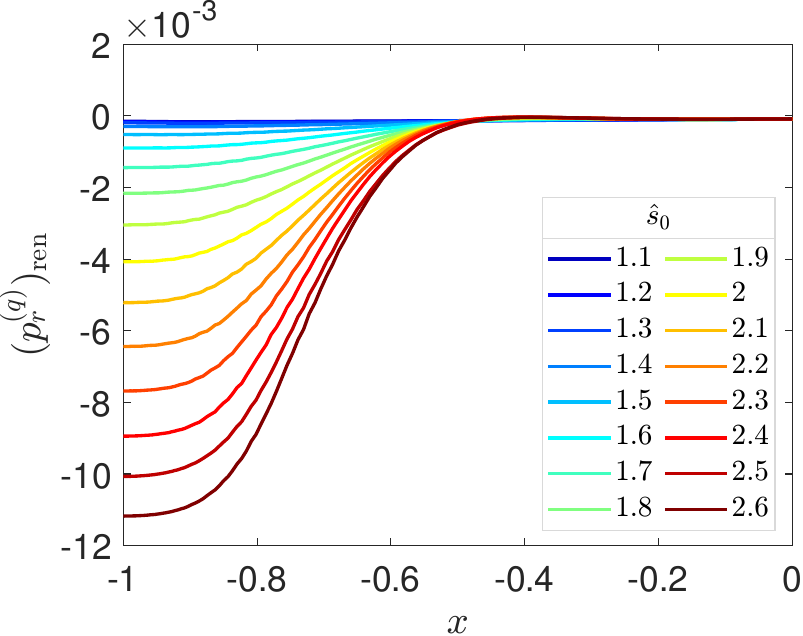}
    \caption{Renormalized quantum energy density and radial pressure.}
    \label{fig:renStress}
\end{figure}

The renormalized quantum stress tensor in \eref{eq:renDensity} is presented in \fig{fig:renStress}. The energy density is positive over most of the region and its center value becomes larger as the curvature grows. A region of small negative energy density is around the central region near $x=-0.6$, which we interpret as an indication that the classical boson star is close to, but not exactly, a self-consistent solution of the semiclassical Einstein equation once quantum backreaction on the metric is taken into account. In contrast, the radial pressure is negative, with a magnitude that increases for larger curvature. This negative pressure suggests that the quantum backreaction tends to reduce the stability of the spacetime, and the classical configuration would be modified to reach an equilibrium state. 

\begin{figure}[tbp]
    \centering
    \includegraphics[width=0.55\linewidth]{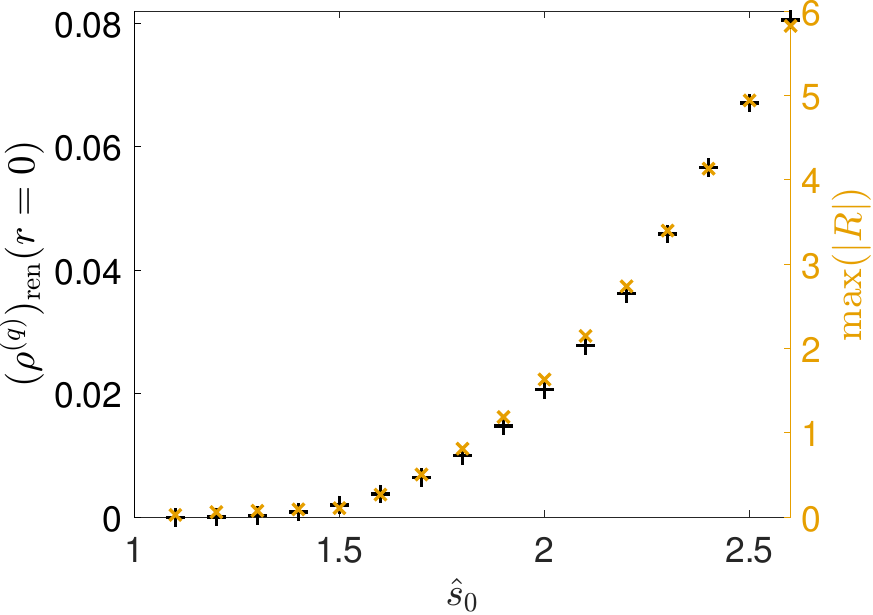}
    \caption{Renormalized quantum energy density in the center and the maximum of the Ricci scalar.}
    \label{fig:ricciInduceStress}
\end{figure}

Using the central value of the renormalized quantum energy density as a measure of the magnitude of quantum effects, we can investigate the key influencing factors. \fig{fig:ricciInduceStress} shows the renormalized quantum energy in the center and the maximum absolute value of the Ricci scalar as functions of $\h{s}_0$. Both quantities increase monotonically as $\h{s}_0$ and exhibit very similar growth patterns. This strong correlation supports the conclusion that spacetime curvature is the primary factor that controls the size of quantum effects. 

\begin{figure}[tbp]
    \centering
    \includegraphics[width=0.48\linewidth]{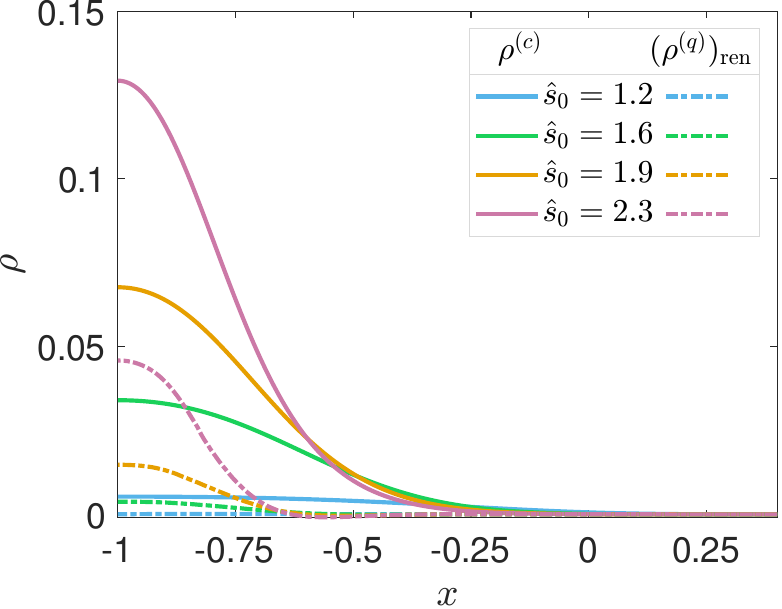}
    \hfill
    \includegraphics[width=0.48\linewidth]{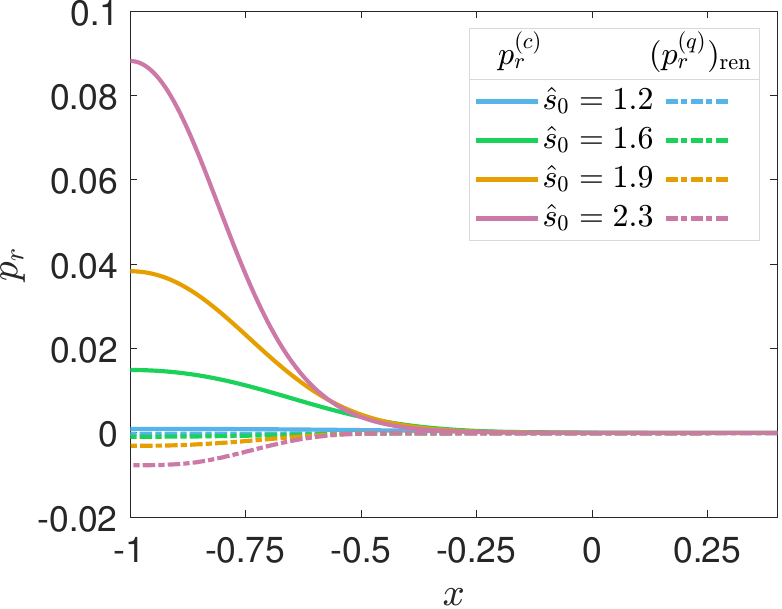}
    \caption{Classical part and quantum fluctuation of the energy density and radial pressure.}
    \label{fig:compareCQ}
\end{figure}

\fig{fig:compareCQ} compares the classical part and the quantum fluctuation of the stress tensor, assuming the Planck mass $(m_P)_{\ren}=1$. Both the classical part and quantum fluctuation of the energy density are positive, while the quantum fluctuation of the radial pressure is negative, opposite to the classical part. For small curvature, the quantum fluctuation is much smaller than the classical part. However, as the curvature increases, it becomes comparable in magnitude. This demonstrates that even in a coherent state, the quantum state most closely resembling to classical physics, quantum effects can become significant in strongly curved spacetimes.

\begin{figure}[tbp]
    \centering
    \includegraphics[width=0.5\linewidth]{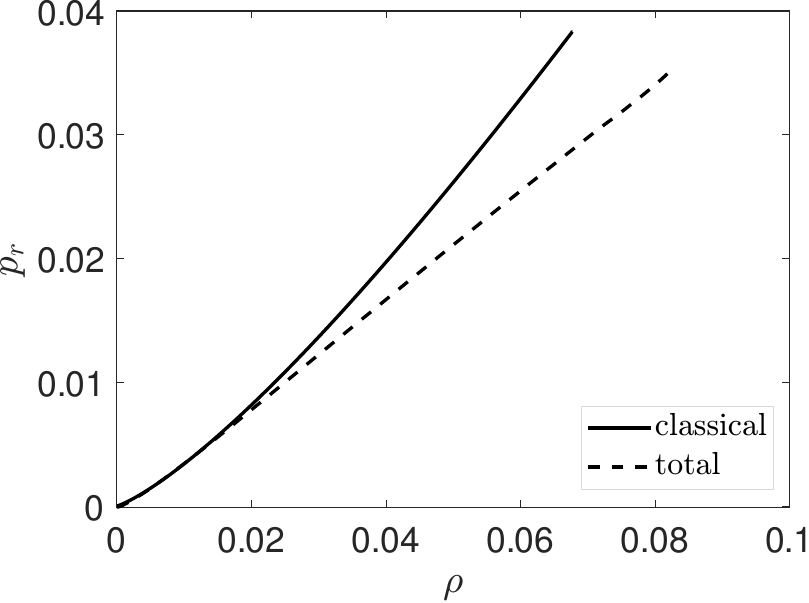}
    \caption{Relation between radial pressure and energy density at different radial positions inside the boson star with $\h{s}_0=1.9$, for the classical part and the total part of the stress tensor. Both quantities are close to zero near the surface and increase towards the center, where they reach their maximum values.}
    \label{fig:polytropicEoS}
\end{figure}

Focusing on the radial pressure and the energy density of a single boson star, \fig{fig:polytropicEoS} shows their relation at different radial positions inside the boson star with $\h{s}_0=1.9$, whose radius is defined in \eref{eq:BSradius}. For the classical part, the radial pressure increases with the energy density, both vanishing at the surface and reaching maxima in the center. A numerical fit shows that this relation is well described by a polytropic form $p_r^{(\text{c})}=1.16(\rho^{(\text{c})}){}^{1.26}$. When quantum fluctuations are included, the relation between the total radial pressure and total energy density becomes closer to linear, with a fit $(p_r)_\ren=0.5(\rho_\ren){}^{1.06}$, indicating a reduced exponent. Similar behaviour is observed for other solutions; although the numerical coefficients vary, the exponent decreases once quantum effects are included. This suggests that quantum effects effectively soften the equation of state, reducing the pressure for a given energy density and potentially leading to less massive configurations. However, since backreaction has not yet been incorporated into the semiclassical Einstein equation, the properties of fully self-consistent solutions remain to be determined. 

\begin{figure}[tbp]
    \centering
    \includegraphics[width=0.55\linewidth]{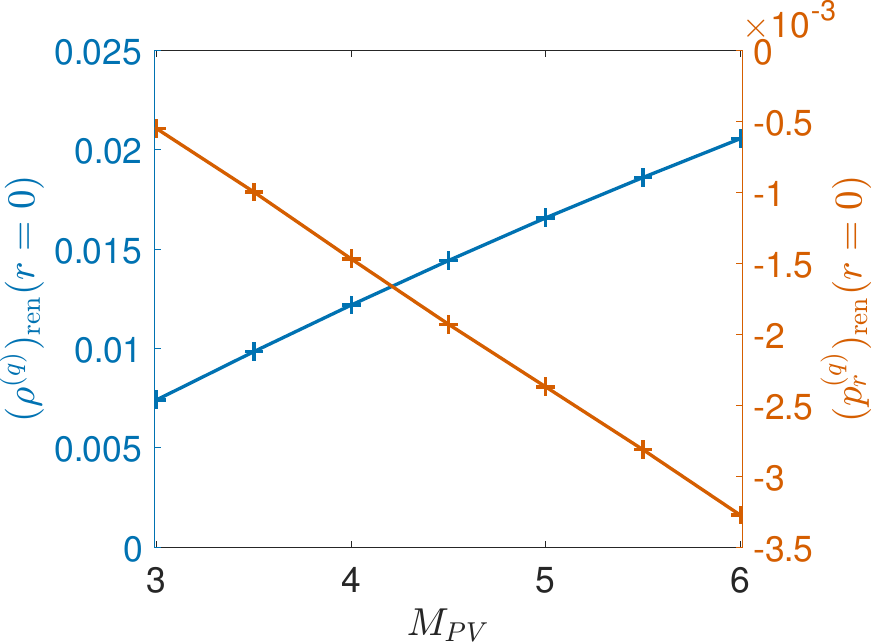}
    \caption{Dependence of the renormalized quantum energy density and radial pressure in the center on the Pauli-Villars mass for the maximum compactness boson star with $\h{s}_0=1.92$.}
    \label{fig:PVmassDependence}
\end{figure}

Lastly, we comment on the dependence of the results on the Pauli-Villars mass $M_{PV}$. Increasing $M_{PV}$ enhances both the bare stress tensor and the counterterms in magnitude, and the renormalized quantum stress tensor is also larger. The dependence of the central energy density and radial pressure is shown in \fig{fig:PVmassDependence}, exhibiting approximately linear behaviour with slight curvature. Since only the leading quartic and quadratic divergences are subtracted, we do not expect the results converge in the large mass limit. It is not clear whether the data is a part of a logarithmic dependence, because larger masses are beyond our numerical capability at the current stage. Although the logarithmic correction is subdominant, including it may improve convergence, which we will investigate in future work \cite{Saffin:2026}. 

Nevertheless, the qualitative features remain unchanged for different $M_{PV}$; the renormalized quantum energy density is mostly positive, and increases with curvature, while the renormalized radial pressure is negative, and decreases further as curvature grows. The small negative energy density region around the boson star persists. In the limit $x\to1$, both $(\ri^{\text{(q)}})_{\ren}$ and $(p_r^{\text{(q)}})_{\ren}$ vanish no matter how large $M_{PV}$ is, consistent with an asymptotically flat spacetime.

\section{Summary and outlook}
\label{sec:summary}

In this work, we compute quantum scalar fields and their stress tensor in coherent states in boson star spacetimes, where the metric is treated classically but matter fields are quantized. Using the Pauli-Villars method, diffeomorphism invariance is manifestly preserved and the regularization applies to each set of momentum modes, suitable for numerical calculation. The quantum mode functions are then evaluated using the spectral method, providing high numerical accuracy. The stress tensor in coherent states is obtained by summing over these modes and shown to converge as the number of included modes increases. A final extrapolation and averaging are used to reach the infinite-mode limit. 

The spectrum of the mode functions contains both bound states and scattering states, with frequencies that scale linearly at large momentum. Contributions to the stress tensor from quantum fluctuations are obtained, which cannot be simply neglected in curved spacetime. The renormalized quantum energy density is predominantly positive, while the radial pressure exhibits negative contribution, with both effects becoming larger near the center of the boson star. We identify that the main source for large quantum effects is the strong spacetime curvature, not relevant directly to the boson star's mass or compactness. Consequently, sizable quantum effects may occur even in systems without superheavy objects. Furthermore, by comparing the quantum stress tensor with the classical counterpart, we find that they can become comparable in the strong curvature cases. 

Our results indicate that classical boson stars can receive substantial quantum corrections. To identify gravitating equilibrium objects in semiclassical gravity, the backreaction of the quantum stress tensor to the spacetime must be included. Our method has the potential to find these solutions through an iteration procedure. This means that we insert the obtained quantum stress tensor into the semiclassical Einstein equation, and solve the metric functions and the classical fields using the Newton-Raphson method, like what we do to find the classical boson star solutions. The resulting metric would then be used to compute new mode functions and stress tensor. We are currently investigating whether repeating this process yields a self-consistent semiclassical boson star solution. 

Such semiclassical boson star solutions may display many intriguing properties. Previous studies suggest that quantum corrections can increase compactness of gravitational objects \cite{Arrechea:2021pvg,Arrechea:2021xkp,Reyes:2023fde,Arrechea:2023oax}, in some cases even exceeding the Buchdahl limit. Whether similar behaviour occurs for boson stars remains an interesting open question, and may have implications in the quest for black hole mimickers. Work in this direction is currently underway. 

Finally, our method can be readily extended to other compact objects, especially those lacking analytical solutions. Although in this paper some boson stars we study have $\h{s}_0$ larger than that of the maximum mass solution, which are dynamically unstable \cite{Seidel:1990jh}, many other compact objects can reach strong-curvature regimes while remaining stable. Examples include scalar boson stars with a self-interaction potential \cite{Colpi:1986ye,Friedberg:1986tq} and boson stars composed of vector fields \cite{Brito:2015pxa}, both of which can attain higher masses, compactness, and curvature than the free scalar case studied here. In addition, stability is not directly tied to the size of quantum corrections, and thus our results remain broadly applicable. We leave the exploration of these generalizations to future work. 

\section*{Data Access Statement}

All data created during this research are openly available from the University of Nottingham data repository at \cite{nottData}. 
%The data that support the findings of this article are openly available [].

\acknowledgments

We would like to thank Edmund Copeland, Sam Dolan, Zong-Zhe Du, Steffen Gielen, Oliver Gould, Joonas Hirvonen, Victor Jaramillo, Elizabeth Winstanley and Shuang-Yong Zhou for helpful discussions. We also thank the anonymous referees for their careful reading of this manuscript and helpful comments which have improved the paper. QXX acknowledges support from CSC (File No.~202406340173), and PMS would like to thank the STFC
for partial support under grant number ST/X000672/1. We are grateful for access to the University of Nottingham's Ada HPC service.

~

\appendix

\noindent{\bf APPENDIX}

\section{Pauli-Villars regularization conditions}
\label{sec:PVinFRW}

In this section, we derive the Pauli-Villars regularization conditions in Friedmann-Robertson-Walker (FRW) spacetime, where adiabatic solutions exist. A theory exhibits the same structure of ultraviolet divergences in all spacetimes, since these divergences are only sensitive to small scale physics and all spacetimes are locally Minkowskian. We are therefore able to compute the Pauli-Villars regularization conditions and the counterterms through the adiabatic expansion \cite{Parker:1974qw}, which is among the most efficient methods for implementing regularization and renormalization in curved spacetimes, and then apply them to other spacetimes. The same results can also be obtained using other methods, such as the point-splitting regularization \cite{Christensen:1976vb}, and physical results are irrelevant of what regularization method is used. The calculation is performed up to second order in the adiabatic expansion \cite{Berczi:2024yhb}. We also compare the results with those of Minkowski spacetime. 

The line element of FRW spacetime is 
\begin{equation}
    ds^2=-\d t^2+a^2(t)\di_{ij}\d x^i\d x^j,
\end{equation}
where Cartesian coordinates are used and $a(t)$ denotes the scale factor. The non-vanishing components of the Einstein tensor are
\begin{equation}
    G_{00}=3\f{\dot{a}^2}{a^2},~~G_{ij}=\di_{ij}(-\dot{a}^2-2a\ddot{a}).
\end{equation}
The spacetime can still be foliated by a set of constant-$t$ spacelike hypersurfaces with unit normal vector $n^{\mu}=(1,\bm{0})$, owing to spatial homogeneity and isotropy, where canonical quantization is performed. The matter part is unchanged as it is in the main text, and both the physical fields and the ghost fields satisfy the Klein-Gordon equation. For clarity, we first consider one single field and include all fields when implementing the regularization. 

By requiring that the spacetime approaches Minkowski space $a(t)\to a_0$ as $t\to-\infty$ (or $+\infty$), the field can be expanded as 
\begin{align}
\phi&=\int d^3p\lt(A_{\bm{p}}f_{\bm{p}}+h.c.\rt),~~f_{\bm{p}}=\f{e^{-i\oi_{\bm{p}}t+i\bm{p}\cdot\bm{x}}}{\sqrt{(2\pi a_0)^32\oi_{\bm{p}}}}, 
\end{align}
where $h.c.$ denotes the Hermitian conjugate, and the dispersion relation becomes
\begin{equation}
    \oi_{\bm{p}}^2=\f{\bm{p}^2}{a_0^2}+\mu^2.
\end{equation}
It is straightforward to prove that the orthonormality relations on each hypersurface
\begin{equation}
    (f_{\bm{p}},f_{\bm{q}})=\di^3(\bm{p}-\bm{q}),\quad (f_{\bm{p}},f_{\bm{q}}^*)=0.
\end{equation}
Because the scalar product is conserved in time, these relations remain valid even when the spacetime deviates from Minkowski space \cite{Parker:2009uva}. Consequently, the field expansion remains valid with the same time-independent $A_{\bm{p}}$, the commutation relation is invariant
\begin{equation}
    [A_{\bm{p}},A_{\bm{q}}^{\dag}]=\di^3(\bm{p}-\bm{q}),
\end{equation}
and the adiabatic vacuum $|0_A\rangle$ satisfies $A_{\bm{p}}|0_A\rangle=0$. This is equivalent to the commutation relation between the field and its conjugate momentum
\begin{equation}
    [\phi(t,\bm{x}),\pi(t,\bm{y})]=i\di^3(\bm{x}-\bm{y}).
\end{equation}

At any fixed time, spatial symmetry allows the mode function to be expressed in the following form 
\begin{equation}
    f_{\bm{p}}=u_{\bm{p}}(t)e^{i\bm{p}\cdot\bm{x}}.
\end{equation}
The equation of motion for $u_{\bm{p}}(t)$ is then 
\begin{equation}
    \ddot{u}_{\bm{p}}+3\f{\dot{a}}{a}\dot{u}_{\bm{p}}+\oi_{\bm{p}}^2(t)u_{\bm{p}}=0,
    \label{eq:FRWuEquation}
\end{equation}
where the frequency is defined by 
\begin{equation}
    \oi_{\bm{p}}^2(t)=\f{\bm{p}^2}{a^2(t)}+\mu^2. 
\end{equation}
We suppress the subscript and write $\oi=\oi_{\bm{p}}$ in the following. From the above equation, one can derive the following relations
\begin{equation}
\begin{aligned}
    \oi\dot{\oi} &= -(\oi^2-\mu^2)\f{\dot{a}}{a}, \\
    \oi\ddot{\oi} &= \lt(-\f{\ddot{a}}{a}+3\f{\dot{a}^2}{a^2}\rt)(\oi^2-\mu^2)-\f{\dot{a}^2}{a^2}\f{(\oi^2-\mu^2)^2}{\oi^2}.
\end{aligned}
\end{equation}

If the spacetime evolves slowly, \eref{eq:FRWuEquation} can be solved using the WKB method \cite{Parker:1974qw}
\footnote{An adiabatic parameter $T$ can be introduced to act as a small parameter for the WKB method, with $T=1$ imposed at the end of the calculation. The adiabatic order is therefore equal to the number of time derivatives. }
\begin{equation}
    u_{\bm{p}}(t)=\f{1}{\sqrt{(2\pi a(t))^32W_{\bm{p}}(t)}}e^{-i\int^tW_{\bm{p}}(t')dt'}.
    \label{eq:WKBansatz}
\end{equation}
We also denote $W=W_{\bm{p}}$. $W(t)$ satisfies 
\begin{equation}
    W^2=\oi^2-\f32\f{\ddot{a}}{a}-\f34\f{\dot{a}^2}{a^2}+\f34\f{\dot{W}^2}{W^2}-\f12\f{\ddot{W}}{W},
    \label{eq:WKBequation}
\end{equation}
which can be solved iteratively. At zeroth adiabatic order one has $(W_0{})^2=\oi^2$, where the subscript represents the adiabatic order. Substituting it into the right-hand side of \eref{eq:WKBequation} yields, to the next order, 
\begin{equation}
    W^2\approx (W_0{})^2+(W_2{})^2+\dots = \oi^2-\f{2\oi^2+\mu^2}{2\oi^2}\f{\ddot{a}}{a}-\f{4\oi^4+4\oi^2\mu^2-5m^4}{4\oi^4}\f{\dot{a}^2}{a^2}+\cdots.
    \label{eq:WKBfrequency}
\end{equation}

The non-vanishing vacuum expectation values of the stress-energy tensor can be expressed as 
\begin{equation}
\begin{aligned}
    \ri &= \langle0_A|T_{00}|0_A\rangle = 4\pi\int dp\bm{p}^2\f12\lt(|\dot{u}_{\bm{p}}|^2+\f{\bm{p}^2}{a^2}|u_{\bm{p}}|^2+\mu^2|u_{\bm{p}}|^2\rt), \\
    pa^2 &= \f13\di^{ij}\langle0_A|T_{ij}|0_A\rangle = \f13 4\pi\int dp\bm{p}^2\f12\lt(3a^2|\dot{u}_{\bm{p}}|^2-\bm{p}^2|u_{\bm{p}}|^2-3a^2\mu^2|u_{\bm{p}}|^2\rt). 
\end{aligned}
\end{equation}
Using \eref{eq:WKBansatz}, one finds 
\begin{equation}
\begin{aligned}
    |u|^2&=\f{1}{(2\pi a)^32W}, \\
    \dot{u}&=\lt(-\f{\dot{W}}{2W}-\f{3\dot{a}}{2a}-iW\rt)u, \\
    |\dot{u}|^2&=\lt(\f14\f{\dot{W}^2}{W^2}+\f94\f{\dot{a}^2}{a^2}+\f32\f{\dot{a}}{a}\f{\dot{W}}{W}+W^2\rt)\f{1}{(2\pi a)^32W}.
\end{aligned}
\end{equation}
To the required adiabatic order, 
\begin{align}
    \f{1}{W}\approx (W_0{})^{-1}\lt(1-\f12\f{(W_2{})^2}{(W_0{})^2}\rt),~~
    \f{\dot{W}}{W}\approx\f{\dot{W_0}}{W_0}.
\end{align}
We therefore obtain 
\begin{align}
    \ri&=\f{1}{4\pi^2a^3}\int dp\bm{p}^2\lt(\oi+\f18\f{\dot{a}^2}{a^2}\f{(2\oi^2+\mu^2)^2}{\oi^5}\rt), \\ 
    &=\f{1}{4\pi^2a^3}\int dp\bm{p}^2\lt(\sqrt{\bm{p}^2/a^2+\mu^2}+\f18\f{\dot{a}^2}{a^2}\f{(2\bm{p}^2/a^2+3\mu^2)^2}{(\bm{p}^2/a^2+\mu^2)^{5/2}}\rt), \\
    pa^2&=\f13\f{1}{4\pi^2a^3}\int dp\f{\bm{p}^2a^2}{\oi}\bigg(\oi^2-\mu^2 \notag \\ 
    & ~~~~~~~~~~~~~~~~~ +\f18\f{\dot{a}^2}{a^2}\f{(2\oi^2+\mu^2)(2\oi^4-\oi^2\mu^2+5m^4)}{\oi^6}-\f{(2\oi^2+\mu^2)^2}{4\oi^4}\f{\ddot{a}}{a}\bigg), \\
    &=\f13\f{1}{4\pi^2a^3}\int dp\f{\bm{p}^2a^2}{\oi}\bigg(\bm{p}^2/a^2 \notag \\ 
    & ~~~~~~~~~~~~~~~~~ +\f18\f{\dot{a}^2}{a^2}\f{(2\bm{p}^2/a^2+3\mu^2)(2\bm{p}^4/a^4+3(\bm{p}^2/a^2)\mu^2+6m^4)}{(\bm{p}^2/a^2+\mu^2)^3}-\f{(2\bm{p}^2+3\mu^2)^2}{4(\bm{p}^2/a^2+\mu^2)^2}\f{\ddot{a}}{a}\bigg).
\end{align}
The resulting momentum integrals are divergent. 

To implement regularization, we now includes all fields, including the Pauli-Villars ghost fields. Introducing an ultraviolet cutoff $M$ for the momentum integration\footnotemark{}, one obtains 
\begin{align}
    4\pi^2\ri &= \sum_n(-1)^{(n)}\f14\bigg\{ M^4+M^2(m\ind{n})^2+\f{(m\ind{n})^4}{8}-\f{(m\ind{n})^4}{4}\ln\f{4M^2}{(m\ind{n})^2}+O\lt(\f{(m\ind{n})^6}{M^2}\rt) \notag \\ 
    & ~~~~~~~~~~~~~~ +\f{\dot{a}^2}{a^2}\lt[M^2-\f{4(m\ind{n})^2}{3}+\f{(m\ind{n})^2}{2}\ln\f{4M^2}{(m\ind{n})^2}+O\lt(\f{(m\ind{n})^4}{M^2}\rt)\rt] \bigg\}, \\
    4\pi^2pa^2 &= \f13\sum_n(-1)^{(n)}\f14\bigg\{ a^2\lt[M^4-M^2(m\ind{n})^2-\f{7(m\ind{n})^4}{8}+\f{3(m\ind{n})^4}{4}\ln\f{4M^2}{(m\ind{n})^2}+O\lt(\f{(m\ind{n})^6}{M^2}\rt)\rt] \notag \\ 
    & ~~~~~~~~~~~~~~~~~ +\dot{a}^2\lt[M^2+\f{7(m\ind{n})^2}{3}-\f{(m\ind{n})^4}{2}\ln\f{4M^2}{(m\ind{n})^2}+O\lt(\f{(m\ind{n})^4}{M^2}\rt)\rt] \notag \\
    & ~~~~~~~~~~~~~~~~~ +2a\ddot{a}\lt[-M^2+\f{4(m\ind{n})^2}{3}-\f{(m\ind{n})^4}{2}\ln\f{4M^2}{(m\ind{n})^2}+O\lt(\f{(m\ind{n})^4}{M^2}\rt)\rt] \bigg\}.
\end{align}
Cancellation of the quartic, quadratic, and logarithmic divergences requires the following conditions
\begin{equation}
    \sum_n(-1)^{(n)}=0,~~\sum_n (-1)^{(n)}(m\ind{n})^2=0,~~\sum_n (-1)^{(n)}(m\ind{n})^4=0.
    \label{eq:PVregConditions_copy}
\end{equation}
\footnotetext{Alternatively, one can perform the summation over fields before the momentum integration, in which case it is not necessary to introduce the cutoff $M$; the regularization is thus cutoff-independent \cite{Visser:2016mtr} and preserves diffeomorphism invariance. } 

Because the Pauli-Villars mass should be sent to infinity, the remaining terms still diverge. Following the spirit of adiabatic subtraction \cite{Parker:1974qw}, if divergences appear at a given adiabatic order, the whole order should be subtracted. Accordingly, we define 
\begin{align}
    \langle T_{00}\rangle_{\text{fin}}&=\langle T_{00}\rangle-g_{00}\di\Li-G_{00}\di\mbar^2, \\
    \di^{ij}\langle T_{ij}\rangle_{\text{fin}}&=\di^{ij}\langle T_{ij}\rangle-g_i{}^i\di\Li-G_i{}^i\di\mbar^2,
\end{align}
where the divergent terms are given by 
\begin{equation}
\begin{aligned}
    \di\Li&=\f{-1}{32\pi^2}\sum_{n}(-1)^{(n)}(m\ind{n})^4\ln m\ind{n}, \\
    \di\mbar^2&=\f{-1}{48\pi^2}\sum_{n}(-1)^{(n)}(m\ind{n})^2\ln m\ind{n}.
    \label{eq:counterterm_copy}
\end{aligned}
\end{equation}
They are identified as counterterms and absorbed into the bare cosmological constant and the bare Planck mass respectively during renormalization. 

These results \eref{eq:PVregConditions_copy} and \eref{eq:counterterm_copy} are universal and can be applied directly to other spacetimes. If higher adiabatic orders are included, additional coupling constants associated with squared curvature terms are required to absorb the corresponding divergences. Since the renormalization scale is chosen where such contributions are vanishing and their scale dependence is only logarithmic and therefore subleading compared to those in $\di\Li,\di\mbar^2$, they are not considered further here \cite{Tranberg:2008ae}. 

Lastly, we consider the case of Minkowski spacetime. The above calculation can be repeated straightforwardly by fixing $a(t)=1$. Any terms containing $\dot{a}$ then vanish, and one can not obtain the expression for $\di\mbar^2$. This indicates that in Minkowski spacetime, only the leading order subtraction is required, which corresponds to a renormalization of cosmological constant. 

\section{Spectral method}
\label{sec:spectral}

In this section, we provide a brief introduction to the spectral method, and exhibit the spectral differentiation matrix explicitly \cite{trefethen2000spectral}. Latin letters are used to denote lattice indices. 

In the spectral method, a function $v(x)$ is expanded using a set of basis functions on a lattice $\{x_j\},j=0,1,\dots,N$, which possesses $N+1$ grid points. In particular, we employ the cardinal function defined by 
\begin{equation}
    p_j(x)=\f{1}{a_j}\prod_{\substack{k=0\\k\ne j}}^N(x-x_k),~~a_j=\prod_{\substack{k=0\\k\ne j}}^N(x_j-x_k).
\end{equation}
This is a polynomial with highest order term $x^N$ that takes the value $1$ at $x_j$ and $0$ at all other points, analogous to a discrete delta function on this lattice. The expansion is then given by 
\begin{equation}
    v(x)\simeq\sum_{j=0}^N v(x_j)p_j(x),
    \label{eq:spectralExpansion}
\end{equation}
which exactly matches the function $v$ at each grid point. In other words, it interpolates $v$ on the lattice. 

A single derivative is computed by differentiating \eref{eq:spectralExpansion} 
\begin{equation}
    v'(x)\simeq\sum_{j=0}^N v(x_j)p_j'(x),
\end{equation}
and the error at the grid points exponentially decreases as the number of grid points increases, provided that the function $v$ is sufficiently smooth. We define the spectral differentiation matrix $D$, whose elements are given by the derivative of $p_j$ at point $x_i$
\begin{equation}
\begin{aligned}
    & D_{ij}=p_j'(x_i)=\f{a_i}{a_j(x_i-x_j)},~~i\ne j,
    \\
    & D_{jj}=p_j'(x_j)=\sum_{\substack{k=0\\k\ne j}}^N(x_j-x_k)^{-1}.
\end{aligned}
\label{eq:spectralMatrix}
\end{equation}
In matrix form, the derivative can be written as 
\begin{equation}
    v'_i\simeq \sum_{j=0}^ND_{ij}v_j,
\end{equation}
where $v_j$ and $v'_j$ denote the values of $v$ and its derivative $v'$ at the point $x_j$. Higher-order derivatives can be obtained straightforwardly by multiplying more matrices, for example, 
\begin{equation}
    v''_i\simeq \sum_{j,k=0}^ND_{ij}D_{jk}v_k.
\end{equation}

For a non-periodic and bounded interval, as the one in this paper, Chebyshev points provide a particularly suitable choice of lattice, 
\begin{equation}
    x_j = \cos(j\f{\pi}{N}).
    \label{eq:chebPoints}
\end{equation}
These points cluster near the boundaries of the interval, and coincide with extrema of the Chebyshev polynomial $T_N(x)$, except the endpoints. One advantage of using Chebyshev points rather than a set of evenly spaced points is the avoidance of the Runge phenomenon, where the interpolants oscillate rapidly near the boundaries, leading to poor convergence. Note that \eref{eq:chebPoints} implies that $x\in[-1,1]$, so for a generic interval, the coordinate should be transformed into this range. 

With Chebyshev points, the differentiation matrix \eref{eq:spectralMatrix} can be written in a simplified form 
\begin{equation}
\begin{aligned}
& D_{00}=\f{2N^2+1}6,~~D_{NN}=-\f{2N^2+1}6, \\
& D_{jj}=\f{-x_j}{2(1-x_j^2)},~~j=1,\dots,N-1, \\
& D_{ij}=\f{c_i}{c_j}\f{(-1)^{i+j}}{(x_i-x_j)},~~i\ne j,
\end{aligned}
    \label{eq:spectralMatrix_cheb}
\end{equation}
where 
\begin{equation}
    c_i=
    \begin{cases}
        2,~~~~i=0\text{ or }N, \\
        1,~~~~\text{otherwise}.
    \end{cases}
\end{equation}
In practical numerical calculations, the diagonal elements are obtained using the off-diagonal elements in the same column
\begin{equation}
    D_{ii}=-\sum_{\substack{j=0\\j\ne i}}^ND_{ij},
\end{equation}
which produces numerically more stable results. Boundary conditions are imposed by replacing the first and/or last rows of the matrix, which denote the right and left boundaries respectively, with the exact forms of the boundary conditions.

% \paragraph{Note added.} 

% Bibliography

%% [A] Recommended: using JHEP.bst file
\bibliographystyle{JHEP}
\bibliography{biblio.bib}

%% or
%% [B] Manual formatting (see below)

% \begin{thebibliography}{99}

% \bibitem{a}
% Author,
% \emph{Title},
% \emph{J. Abbrev.} {\bf vol} (year) pg.

% \bibitem{b}
% Author,
% \emph{Title},
% arxiv:1234.5678.

% \bibitem{c}
% Author,
% \emph{Title},
% Publisher (year).

% \end{thebibliography}

% \printindex
\end{document}